\documentclass[twocolumn]{aastex631}

    \usepackage[flushleft]{threeparttable}

\begin{document}

\title{Determining the methanol deuteration in the disk around V883 Orionis with laboratory measured spectroscopy}

\author[0000-0003-3721-374X]{Shaoshan Zeng}
\affiliation{Star and Planet Formation Laboratory, Pioneering Research Institute, RIKEN, 2-1 Hirosawa, Wako, Saitama, 351-0198, Japan}

\author{Jae-Hong Jeong}
\affiliation{Department of Physics and Astronomy, Seoul National University, 1 Gwanak-ro, Gwanak-gu, Seoul 08826, Republic of Korea}

\author{Takahiro Oyama}
\affiliation{Star and Planet Formation Laboratory, Cluster for Pioneering Research, RIKEN, 2-1 Hirosawa, Wako, Saitama, 351-0198, Japan}

\author{Jeong-Eun Lee}
\affiliation{Department of Physics and Astronomy, Seoul National University, 1 Gwanak-ro, Gwanak-gu, Seoul 08826, Republic of Korea}

\author{Yao-Lun Yang}
\affiliation{Star and Planet Formation Laboratory, Cluster for Pioneering Research, RIKEN, 2-1 Hirosawa, Wako, Saitama, 351-0198, Japan}

\author{Nami Sakai}
\affiliation{Star and Planet Formation Laboratory, Cluster for Pioneering Research, RIKEN, 2-1 Hirosawa, Wako, Saitama, 351-0198, Japan}



\begin{abstract}

Deuterium fractionation, as studied through mono-deuterated methanol, is frequently used as a diagnostic tool to trace the physical conditions and chemical evolution of interstellar sources. This study investigates methanol deuteration in the disk around V883 Ori, utilising recent laboratory spectroscopic data for CH$_2$DOH and CH$_3$OD along with ALMA observations. The derived column densities for CH$_2$DOH and CH$_3$OD are (5.14$\pm$0.08) $\times $10$^{16}$ cm$^{-2}$ and (4.22$\pm$0.06) $\times$ 10$^{16}$ cm$^{-2}$, respectively. The analysis demonstrates the influence of spectroscopic data on determining molecular column density, excitation temperature, and, most importantly, the inferred D/H ratio. The D/H ratio for CH$_2$DOH is calculated to be (7.3$\pm$1.5) $\times$ 10$^{-3}$ after applying a statistical correction, whilst the D/H ratio for CH$_3$OD is (1.79$\pm$0.36) $\times$ 10$^{-2}$. The discovery of an unexpectedly low CH$_2$DOH/CH$_3$OD ratio (1.22$\pm$0.02) in V883 Ori, however, raises further questions about the synthesis and chemical processes involved in CH$_3$OD formation. Overall, this study underscores the importance of accurate spectroscopic data for studies of isotopic fractionation and provides new insights into methanol deuteration chemistry in star-forming regions. Future research, combining updated spectroscopy and chemical modelling, will help further constrain these processes across different masses and evolutionary stages.


\end{abstract}

\section{Introduction} \label{sec:intro}
Deuterium fractionation, the enhancement of the D/H ratio relative to its elemental abundance, has been widely used as a powerful diagnostic tool for characterising the physical parameters and chemical evolution of interstellar sources \citep{Ceccarelli2014,Nomura2022}. Whilst deuterated species are readily observed, our understanding of the deuteration fracitonation process have not been fully established. Observations of molecules in various astrophysical regions not only reveal molecular D/H ratios orders of magnitude higher than the elemental value of $\sim$(1.5$-$2.0) $\times$ 10$^{-5}$ \citep{Linsky2003, Prodanovic2010} but also demonstrate significant variations among different molecular species. 

In general, deuterated species are thought to form efficiently in the cold pre-stellar phase due to the low temperature \citep[e.g.,][]{Caselli2012, Ceccarelli2014}, leading to a significant enhancement of the D/H ratio which can be preserved to later stage of star formation. Among all deuterated molecular tracers, methanol (CH$_3$OH) is one of the molecules that exhibits a large enhancement in the D/H ratio. Whilst previous theoretical and laboratory work showed that the gas-phase formation route is insufficient to account for the observed abundances of methanol \citep{Geppert2005,Garrod2006}, CH$_3$OH has been demonstrated to form efficiently on grain surfaces through the hydrogenation of CO \citep{Watanabe2002,Fuchs2009} and the radical-molecule route between CH$_3$O and H$_2$CO \citep{Simons2020, Santos2022}. Presumably, the deuterated isotopologues of CH$_3$OH observed in protostellar regions are formed in the similar formation pathway i.e. successive addition on CO with one of the H atoms being replaced by D, during the cold pre-stellar phases. These deuterated isotopologues along with the main isotopologue are preserved in icy grain mantles and are subsequently released into the gas phase at the later stage, as a consequence of ice mantle sublimation due to protostellar heating. Therefore, measuring the abundance of deuterated methanol enables the study of deuteration fractionation on the icy surfaces of dust grains, providing access to physical conditions such as temperature and density during the early stages of star formation \citep{Taquet2012a, Taquet2013, Taquet2019, Lee2015, Bogelund2018}. In addition, methanol is considered as an important precursor to more complex organic molecules (COMs). Comparing the D/H ratio of methanol to that of other COMs can be key to understanding the inheritance of interstellar molecules.

CH$_2$DOH and CH$_3$OD are singly deuterated forms of methanol commonly used to derive the D/H ratio, as they have been detected across a wide range of masses and evolutionary stages. In low- and high-mass prestellar and protostellar sources, as well as in comets, the CH$_2$DOH/CH$_3$OH ratio is found to span a wider range, from 10$^{-4}$ to 10$^{-1}$ \citep[e.g.][]{Bianchi2017,Bogelund2018,Jorgensen2018,Ospina-Zamudio2019,Ambrose2021,vanGelder2022,Drozdovskaya2021}, whereas the CH$_3$OD/CH$_3$OH ratio spans within two orders of magnitude, from 10$^{-3}$ to 10$^{-1}$ \citep[e.g.][]{Parise2006,Bogelund2018,Taquet2019}. The difference between low-mass and high-mass sources may be due to either the temperature or the timescale of their pre-stellar phases, as suggested by \citet{vanGelder2022}. Besides, the derived CH$_2$DOH/CH$_3$OD ratio generally exhibits a deviation from its statistical value of 3. In theory, if the addition of D atom on both methyl group ($-$CH$_3$) and hydroxyl group ($-$OH) is assumed equally efficient, a statistical weighting of 3 needs to be taken into account since there are three hydrogen on methyl group compared to hydroxyl group \citep{Charnley1997}. Towards low-mass pre-stellar and protostellar sources, CH$_2$DOH/CH$_3$OD ratio is often larger than 3 and can be up to 20 \citep{Parise2002,Parise2006,Ratajczak2011}, which is in contrast to the ratio being smaller than 3 in most of the studies towards intermediate- and high-mass sources \citep{Belloche2016,Bianchi2017,Bogelund2018}. Numerous chemical models, theorectical studies, and laboratory experiments have been dedicated to elucidating the deviation from the statistical value of three \citep[e.g.,][]{Charnley1997,Osamura2004,Nagaoka2005,Ratajczak2009,Taquet2014,Faure2015,Taquet2019,Kulterer2022,Wilkins2022}.

In order to develop a more comprehensive understanding of the deuteration mechanism of methanol, one crucial step is to accurately evaluate the column density of both CH$_2$DOH and CH$_3$OD. And this relies heavily on well-understood spectroscopy, for which the transition frequency and line intensity are known precisely. As recently studied by \citet{Oyama2023}, significant systematic discrepancies are uncovered between laboratory measurement of CH$_2$DOH spectroscopy and theoretical calculations listed in Jet Propulsion Laboratory\footnote{\url{https://spec.jpl.nasa.gov/ftp/pub/catalog/catdir.html}} \citep[JPL,][]{Pickett1998}, which is widely used in previous observational studies. The most notable difference is found in the value of intrinsic line intensity, \textit{S}$\mu^2$, where \textit{S} is the line strength and $\mu^2$ is the dipole moment which is expected to consequently affect the derivation of column density of CH$_2$DOH. Additionally, the latest spectroscopic study of CH$_3$OD, with new measurements and more refined theoretical model, has been made available by \citet{Ilyushin2024}. In this work, methanol deuterium fractionation in the protoplanetary disk around the young outbursting star V883 Ori is determined using an ALMA line survey and the most up-to-date spectroscopic data. An outline of the spectroscopic and observational data is provided in Sect. \ref{sec:methods}, and the analysis of the spectra, including the derivation of column density and excitation temperature, is presented in Sec. \ref{sec:res}. The discussion and conclusions of the study are given in Sect. \ref{sec:dis} and Sect. \ref{sec:con}, respectively.

\section{Data} \label{sec:methods}

\subsection{Observations} \label{subsec:obs}
V883 Ori was observed by the {\it ALMA Spectral Survey of An eruptive Young star, V883 Ori (ASSAY)} project \citep{Lee2024} in ALMA Cycle 7 (project id: 2019.1.00377.S, PI: Jeong-Eun Lee). The pointing center was set to be a central region of V883 Ori at $\alpha_{J2000}$ = 05$^{h}$38$^{m}$18$^{s}$.100; $\delta_{J2000}$ = -07$^{\circ}$02$\arcmin$25$\arcsec$.980. The spectral survey observation was carried out by three science goals, which consist of 15 tunings with 4 spectral windows (SPWs) per tuning, between May and November 2021 using 12 m array dishes. It is an unbiased spectral survey covering from 221 to 275 GHz in Band 6 with the angular resolution of 0.15$\arcsec$$-$0.2$\arcsec$ (60$-$80 au) and the spectral resolution of 488.281 kHz (0.66$-$0.53 km s$^{-1}$). The full observing log and detailed calibration process are provided in \citet{Lee2024}, while the spectral extraction method and extracted spectra appear in \citet{Yun2023} and \citet{Jeong2025}, respectively. The resulting average RMS noise level of 60 SPWs is 2.4 mJy beam$^{-1}$.

\begin{deluxetable*}{ccrlcccc}
  \tabletypesize{\scriptsize}
  \tablecolumns{8}
  \tablewidth{\textwidth}
  \tablecaption{Spectroscopic information from both JPL theoretical calculation \citep{Pickett1998,Pearson2012} and SUMIRE laboratory measurement\citep{Oyama2023} of the unblended transitions of CH$_2$DOH detected towards V883 Ori. \label{tab:lines}}
  \tablehead{\colhead{Freq.(JPL)$^{(a)}$} & \colhead{Freq.(SUMIRE)$^{(b)}$} & \multicolumn{2}{c}{Transition$^{(c)}$} & \colhead{$E\rm_{up}^{(d)}$} & \colhead{S${\mu^2} (\rm {JPL})^{(e)}$} & \colhead{S${\mu^2} (\rm {SUMIRE})^{(f)}$} & \colhead{$\log I\rm(SUMIRE)^{(g)}$} \\
  \colhead{(MHz)} & \colhead{(MHz)} & \colhead{\textit{J$^{\prime}$K$_a^{\prime}$K$_c^{\prime}$v$\rm _t^{\prime}$}} & \colhead{\textit{J$^{\prime\prime}$K$_a^{\prime\prime}$K$_c^{\prime\prime}$v$\rm _t^{\prime\prime}$}} & \colhead{(K)}  & \colhead{(D$^2$)} & \colhead{(D$^2$)} &  \colhead{(nm$^2$MHz)}}
  \startdata
    \hline
    \multicolumn{8}{c}{\textit{a}-type} \\
    \hline
    221273.004 & 221273.060 & 5 1 5	2 &	4 1	4 2 & 54.7 & 3.66 & 4.51  & -4.1282 \\
    222741.460 & 222741.492 & 5 0 5 1 & 4 0 4 1 & 45.6 & 3.69 & 4.65 & -4.096 \\
    223102.561 & 223102.485 & 5 4 1	1 &	4 4	0 1 & 104.2 & 1.36 & 1.58  & -4.6482 \\
    223102.562 & 223102.485 & 5 4 2	1 &	4 4	1 1 & 104.2 & 1.36 & 1.58  & -4.6482 \\
    223107.081 & 223107.243 & 5 0 5	2 &	4 0	4 2 & 50.5 & 3.73 & 4.73 & -4.0941 \\
    223153.515 & 223153.667 & 5 3 3	2 &	4 3	2 2 & 87.4 & 2.46 & 2.80 & -4.3752 \\
    223153.621 & 223153.667 & 5 3 2	2 &	4 3	1 2 & 87.4 & 2.46 & 2.80 & -4.3752 \\
    223194.716 & 223194.743 & 5 2 3	2 &	4 2	2 2 & 67.6 & 3.38 & 3.87 & -4.2058 \\
    223196.401 & 223196.499 & 5 0 5	0 &	4 0	4 0 & 32.2 & 3.04 & 3.60 & -4.1858 \\
    223315.468 & 223315.387 & 5 2 3	1 &	4 2	2 1 & 58.7 & 3.24 & 3.68 & -4.2143 \\
    223422.058 & 223422.011 & 5 2 4	0 &	4 2	3 0 & 48.3 & 2.45 & 2.87 & -4.3068 \\
    224928.016 & 224928.011 & 5 1 4	2 &	4 1	3 2 & 55.3 & 3.67 & 4.32 & -4.1333 \\
    225667.709 & 225667.591 & 5 1 4	1 &	4 1	3 1 & 49.0 & 3.65 & 4.45 & -4.1085 \\
    226818.248 & 226818.357 & 5 1 4	0 &	4 1	3 0 & 36.7 & 2.90 & 3.46 & -4.1956 \\
    \hline
    \multicolumn{8}{c}{\textit{b}-type} \\
    \hline
    221683.458 & 221683.227 & 5 1 5	1 &	4 2	2 0 & 48.2 &	2.58 & 2.94	& -4.3003 \\
    225878.232 & 225878.125 & 3	1 3	2 &	2 0	2 2 & 35.6 &	1.69 & 1.65	& -4.5192 \\
    227113.050 & 227113.407 & 14 2 12 0 & 14 1 13 0 & 243.0	& 18.94	& 12.74	& -3.9270 \\
    237249.907 & 237249.510 & 7	2 5	0 &	7 1	6 0 & 76.4	& 7.84 & 5.89 & -3.9825 \\
    240643.524 & 240643.640 & 7	0 7	0 &	6 1	6 0 & 60.0	& 4.33 & 3.86 & -4.1298 \\
    244841.135 & 244841.017 & 4	2 2	0 &	4 1	3 0 & 37.6	& 2.54 & 2.01 & -4.3657 \\
    247625.746 & 247625.803 & 3	2 1	0 &	3 1	2 0 & 29.0	& 2.36 & 1.88 & -4.3724 \\
    247846.432 & 247846.509 & 9	2 7	2 &	9 1	8 2 & 131.9	& 2.95 & 2.17 & -4.4582 \\
    255647.816 & 255647.980 & 3 2 2 0 & 3 1 3 0 & 29.0 & 2.25 & 1.78 & -4.3681 \\
    257929.473 & 257929.905 & 7	0 7	1 &	6 1	6 1 & 73.3	& 1.13 & 0.90 & -4.7206 \\
    261263.922 & 261263.948 & 11 1 11 0 & 10 0 10 1 & 143.4	& 5.26	& 5.15	& -4.0532 \\
    \hline
    \multicolumn{8}{c}{\textit{c}-type} \\
    \hline
    224928.579 & 224928.626 & 10 0 10 1 &	9 1 8 2 & 130.8 & 1.71 & 4.53	& -4.2221 \\
    241027.458 & 241027.442 & 1 1 0 2 &	2 0 0 0 & 25.0 & 0.63 & 0.75	& -4.7450 \\
    257895.673 & 257895.586 & 4 2	3 1 &	3 1 3 2 & 48.0 & 2.85 & 1.42	& -4.4861 \\
  \enddata
   \tablecomments{$^{(a)}$ Frequency reported in JPL database. $^{(b)}$ Frequency measured by SUMIRE. $^{(c)}$Quantum numbers for the upper ($^{\prime}$) and lower state ($^{\prime\prime}$) of each transitions with J denotes the angular momentum; \textit{K$_a$} and \textit{K$_c$} denotes the projections of \textit{J} along the \textit{a} and \textit{c} inertial axes respectively; torsional sub-states, e$_0$, e$_1$, and o$_1$ are denoted with \textit{v$\rm _t$} = 0, 1, and 2, respectively. $^{(d)}$Upper state energy relative to the ground state. $^{(e)}$Values of S${\mu^2}$ listed in JPL catalogue. $^{(f)}$Values of \textit{S}${\mu^2}$ derived by SUMIRE. $^{(g)}$Base 10 logarithm of the integrated intensity using \textit{S}${\mu^2}$ derived by SUMIRE.
   }
\end{deluxetable*}

\subsection{Spectroscopic Data} \label{subsec:spec}
The spectroscopic data of CH$_2$DOH were obtained from measurement carried out by Spectrometer Using superconductor MIxer REceiver (SUMIRE) in RIKEN in the range of 216 GHz $-$ 264 GHz. Full details of the experimental setup and the spectroscopic analysis is described in \citet{Watanabe2021} and \citet{Oyama2023} respectively. Unlike conventional absorption spectrometers, a ALMA-cartridge-type receiver was employed to detect the emission of the CH$_2$DOH sample gas, meaning that its thermal emission spectrum with accurate transition frequency and absolute intensities can be obtained directly in the laboratory measurements. The Fast Fourier Transform Spectrometers (FFTS) backend is used to provide the frequency resolution and bandwidth of 88.5 kHz (equivalent to a velocity resolution of $\sim$0.1 km s$^{-2}$ in the covered frequency range) and 2.5 GHz respectively. The achieved root-means-square (rms) noise is 19 $-$ 48 mK. The full spectroscopic information of CH$_2$DOH, including values of rest frequencies and experimentally-determined S${\mu^2}$ used in this study, can be found in Table 1 in Appendix B of \citet{Oyama2023}. Based on the experimental results, the partition function of CH$_2$DOH is also re-evaluated by taking into account higher rovibrational levels which are not included in the calculation for the JPL database. The newly computed partition function for CH$_2$DOH is listed in Table \ref{tab:Qrot}. Under the assumption of LTE conditions, the total partition function is represented by the product of the rotation term, Q$\rm _r$(T), the vibration term, Q$\rm _v$(T), and the torsion term, Q$\rm _t$(T) \citep[see equations 6$-$9 in][]{Oyama2023}. Whilst classical approximation is used for the rotation term, the vibration term is calculated by using the fundamental vibration frequencies of CH$_2$DOH \citep{Serrallach1974}, and the torsion term is evaluated by using the energies of torsional sublevels from e$_0$ to o$_7$ calculated by \citet{Mukhopadhyay1997}. In addition, the contribution of nuclear spin to the partition function was ignored due to the fact that the hyperfine splitting is small and negligible in the high-\textit{J} transitions. However, the effect of nuclear spin statistics on the partition function for CH$_2$DOH is not well known, further studies with better modelling and broader frequency coverage of the measured spectrum will be necessary to improved the evaluation.

\section{Analysis and Results} \label{sec:res}

\subsection{Spectra Corrected for Disk Rotation and Line Identification}\label{subsec:line}
To investigate sublimated COMs in the Keplerian disk around V883 Ori, \citet{PCA_2023} developed a first principal component (PC1) filtering method to extract spectra corrected for the disk rotation over the entire COM emission region, which has a radius of $\sim$0.3$\arcsec$, approximately. The PC1 filter was constructed by applying the Principle Component Analysis (PCA) to the cube data of strong and well-isolated COMs lines. As a result, the PC1 filter describes the most common kinematic and spatial distribution of the COMs on the disk. The PC1-filtered spectra over the full frequency range are very powerful for identifying weak lines of low-abundance species because of the significantly improved signal-to-noise ratio of the lines. Refer to the \citet{PCA_2023} for details of the PC1-filtering method.

For line identification, the observational data were imported into the Spectral Line Identification and Modelling (SLIM) tool within \textsc{madcuba} package\footnote{Madrid Data Cube Analysis on ImageJ is a software developed at the Center of Astrobiology (CAB) in Madrid; \url{http://cab.inta-csic.es/madcuba/}}\citep[version 08/31/2023,][]{Martin2019}. Originally, \textsc{madcuba} incorporates the spectroscopic entry from JPL (ID: C033004) for CH$_2$DOH which is mainly based on the study of \citet{Pearson2012}. In addition, the new spectroscopic data from \citet{Oyama2023} (hereafter SUMIRE) was added to \textsc{madcuba} as an additional catalogue entry of CH$_2$DOH. For CH$_3$OD, the spectra data is taken from the Cologne Database for Molecular Spectroscopy\footnote{\url{https://cdms.astro.uni-koeln.de/classic/}} \citep[CDMS, ID: 033511][]{Endres2016} which is based on the work by \citet{Ilyushin2024}. The corresponding partition function for both CH$_2$DOH and CH$_3$OD at a set of defined temperature is also imported into \textsc{madcuba}. The partial derivative of the partition function with respect to the excitation temperature is approximated in \textsc{madcuba} by the linear slope between the two values from the entries surrounding the given excitation temperature. In this way, the same analyse routine can be executed on the observational data for CH$_2$DOH and CH$_3$OD with different spectroscopic entries simultaneously.


\begin{figure*}
\includegraphics[width=\textwidth]{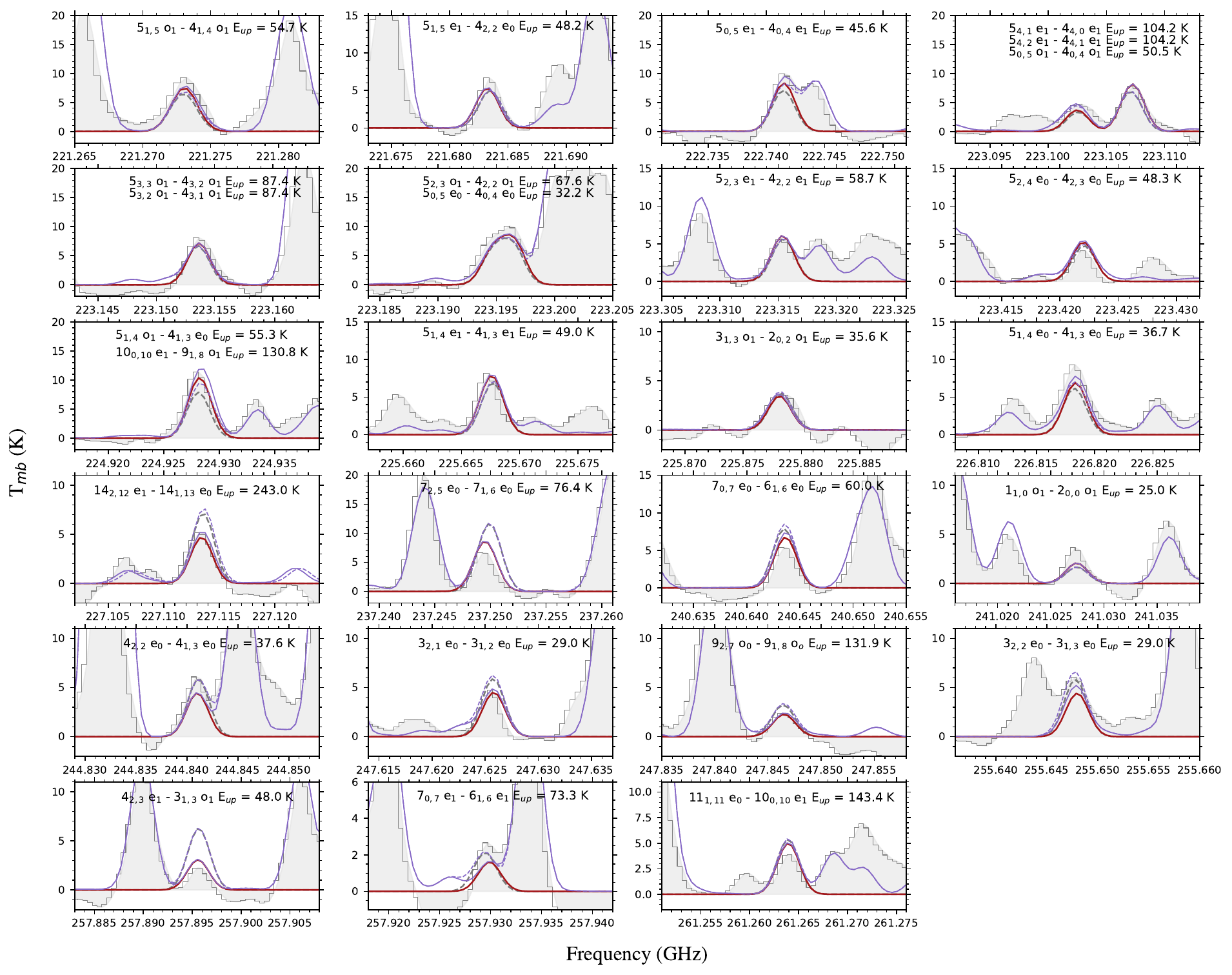}
\centering
\caption{Unblended and optically thin ($\tau \leq$0.3) transitions of CH$_2$DOH detected in V883 Ori. The observed spectra are plotted as grey histograms in the order of increasing frequency. Assuming the same excitation temperature (T${\rm _{ex}}$=120 K), radial velocity ($\rm \nu_{LSR}$), linewidth (FWHM=3.0 km\,s$^{-1}$), and beam-filling factor (0.384$^{\prime\prime}$), the result of the best LTE fit using spectroscopic data from JPL and SUMIRE is shown in grey dashed line and red line respectively. The overall fitting i.e. by including the contribution from all other molecular species detected in V883 Ori in \citet{Jeong2025}, is in purple dashed line (JPL) and purple line (SUMIRE). The quantum numbers and value of E$\rm _{up}$ of each transition are listed on the upper part of each panel.
\label{fig:CH2DOH}}
\end{figure*}

\begin{deluxetable*}{cccc|ccc}
  \tabletypesize{\scriptsize}
  \tablewidth{\textwidth}
  \tablecaption{Derived column densities, excitation temperatures, and resulting CH$_2$DOH/CH$_3$OH ratio of CH$_2$DOH detected towards V883 Ori using different values of \textit{S}$\mu^2$. \label{tab:gaussian-fit-parameters}}
    \tablehead{\colhead{} & \colhead{T$\rm_{ex}$} & \colhead{$\textit{N}$} & \colhead{CH$_2$DOH/CH$_3$OH$^{b}$} & \colhead{T$\rm_{ex}$} & \colhead{$\textit{N}$} & \colhead{CH$_2$DOH/CH$_3$OH$^{b}$} \\
    \colhead{} & \colhead{(K)} & \colhead{($\rm \times$10$^{16}$ cm$^{-2}$)} & \colhead{($\times$10$^{-2}$)} & \colhead{(K)} & \colhead{($\rm \times$10$^{16}$ cm$^{-2}$)} & \colhead{($\times$10$^{-2}$)}}
      \startdata
    \hline
     & \multicolumn{3}{c}{JPL} & \multicolumn{3}{c}{SUMIRE} \\
    \hline
     \textit{a}-type transitions & 103$\pm$10 & 5.6$\pm$0.7 & 2.4$\pm$0.6 & 106$\pm$10 & 4.6$\pm$0.5 & 2.0$\pm$0.5 \\
     \textit{b}-type transitions & 88$\pm$7 & 3.1$\pm$0.3 & 1.3$\pm$0.3 & 99$\pm$8 & 3.8$\pm$0.4 & 1.6$\pm$0.4 \\
    All transitions & 92$\pm$5 & 4.3$\pm$0.3 & 1.8$\pm$0.4 & 102$\pm$6 & 4.3$\pm$0.3 & 1.8$\pm$0.4 \\
    \hline
     \textit{a}-type transitions & 120$^{a}$ & 6.7$\pm$0.1 & 2.8$\pm$0.6 & 120$^{a}$ & 5.3$\pm$0.1 & 2.3$\pm$0.4 \\
     \textit{b}-type transitions & 120$^{a}$ & 4.1$\pm$0.2 & 1.8$\pm$0.4 & 120$^{a}$ & 4.4$\pm$0.2 & 1.9$\pm$0.4\\
    All transitions & 120$^{a}$ & 5.7$\pm$0.1 & 2.4$\pm$0.5 & 120$^{a}$ & 5.1$\pm$0.1 & 2.2$\pm$0.4 \\
    \enddata
  \tablecomments{$^{(a)}$ excitation temperature is fixed to 120 K to be consistent with that derived from the disk model as reported in \citet{Lee2019}. $^{b}$ D/H is derived by adopting N(CH$_3$OH) = (2.35$^{+0.65}_{-0.24}$) $\times$ 10$^{18}$ cm $^{-2}$ \citep{Jeong2025}
   }
   \label{tab:derived_results}
\end{deluxetable*}

\subsection{\texorpdfstring{CH$_2$DOH}{lg}}\label{subsec:ch2doh}
Among a plethora of CH$_2$DOH lines detected in the spectral surveys towards V883 Ori, 28 unblended transitions with optical depth $\tau \leq $0.3 are identified. As listed in Table \ref{tab:lines}, these includes \textit{a}-, \textit{b}-, and \textit{c}-type, transitions: 14 \textit{a}-type transitions, 11 \textit{b}-type transitions, and 3 \textit{c}-type transitions. With all these identified lines, the SLIM tool generates synthetic spectra under the assumption of Local Thermodynamic Equilibrium (LTE) conditions and the automatic fitting routine SLIM-AUTOFIT was used to provide the best non-linear least-squares LTE fit to the line profile using the Levenberg-Marquardt algorithm, which provides the value and uncertainty of the physical parameters \citep[see detailed description in][]{Martin2019}. The free parameters of the model are: molecular column density ($\textit{N}_{\rm tot}$), excitation temperature ($\textit{T}_{\rm ex}$), central radial velocity ($\upsilon_{\rm LSR}$), and full width half maximum (FWHM). Derived physical parameters are provided in Table \ref{tab:derived_results}. Figure \ref{fig:CH2DOH} shows the best LTE fit to the unblended and optically thin transitions of CH$_2$DOH detected in V883 Ori by using the spectroscopic data from JPL and SUMIRE independently. 

By comparing the experimental results to the calculated values listed in the JPL database, \citet{Oyama2023} reported systematic deviations on both frequency and intrinsic intensity of CH$_2$DOH transitions. For the CH$_2$DOH transitions detected in this study, the deviation in line frequency ranges from 5 to 432 kHz (0.007 $-$ 0.5 km\,s$^{-1}$). This deviation is much smaller than the FWHM of the line profile (3 km\,s$^{-1}$), therefore, does not lead to misidentification of the transition or significantly influence the derivation of the CH$_2$DOH column density. However, deviations in other CH$_2$DOH transitions can reach up to $\pm$2 MHz ($\sim$2.2 km\,s$^{-1}$) between the measured and calculated frequencies. Such effects may become more significant depending on the FWHM of the astronomical source and the detected transitions.

To investigate the effect from the deviation of S${\mu^2}$ values between JPL and SUMIRE, the CH$_2$DOH column density is derived by using 1) \textit{a}-type transitions, 2) \textit{b}-type transitions, and 3) all detected transitions. There are not enough detected \textit{c}-type transitions to derive reliable CH$_2$DOH column density separately. To be consistent with the analysis of other molecular species detected in V883 Ori, all free parameters besides column density were fixed to the same value as \citet{Jeong2025} which are $\upsilon$ = 0 km\,s$^{-1}$\footnote{This refers to the offset from the source systemic velocity}, FWHM = 3 km\,s$^{-1}$, beam-filling factor = 0.384 $^{\prime\prime}$, and $\textit{T}_{\rm ex}$ = 120 K.  With only \textit{a}-type transitions, one of the strongest $J = 5 - 4$ series that is often targeted in observations, the best LTE fitting gives \textit{N}$\rm _{JPL}$ = (6.7$\pm$0.1) $\times$ 10$^{16}$ cm$^{-2}$ and \textit{N}$\rm _{SUMIRE}$ = (5.3$\pm$0.1) $\times$ 10$^{16}$ cm$^{-2}$. The $\sim$20$\%$ lower in column density is attributed to the fact that the \textit{S}$\mu^2$ values obtained from SUMIRE were in general \citep[13$\% - $27$\%$;][]{Oyama2023} larger in comparison to JPL. By leaving the $\textit{T}_{\rm ex}$ as a free parameter, the derived $\textit{T}_{\rm ex, JPL}$ is 103$\pm$10 K and $\textit{T}_{\rm ex, SUMIRE}$ is 106$\pm$10 K, which are consistent to each other within uncertainty and comparable to the assumed value of 120 K. The derived column density is \textit{N}$\rm _{JPL}$ = (5.6$\pm$0.7) $\times$ 10$^{16}$ cm$^{-2}$ and \textit{N}$\rm _{SUMIRE}$ = (4.6$\pm$0.5) $\times$ 10$^{16}$ cm$^{-2}$.

For \textit{b}-type transitions, particularly of Q-branch, the \textit{S}$\mu^2$ values obtained from SUMIRE is found to be systematically deviated from the JPL values: the ratio between the SUMIRE and JPL values of \textit{S}$\mu^2$ decrease as \textit{J}increases for the e$_0$ and o$_1$ state whilst the ratio increase as \textit{J} increases for the e$_1$ state \citep{Oyama2023}. However, due to the limited  \textit{b}-type transitions detected in V883 as non-blended lines, extensive analysis of transitions within each torsional state can not be made. Using all the available  \textit{b}-type transitions, the best LTE fitting gives \textit{N}$\rm _{JPL}$ = (4.1$\pm$0.2) $\times$ 10$^{16}$ cm$^{-2}$ and \textit{N}$\rm _{SUMIRE}$ = (4.4$\pm$0.2) $\times$ 10$^{16}$ cm$^{-2}$ when $\textit{T}_{\rm ex}$ is fixed to 120 K. The increase in column density is expected from the general decrease of \textit{S}$\mu^2$ value. By leaving the $\textit{T}_{\rm ex}$ as a free parameter, the derived $\textit{T}_{\rm ex, JPL}$ is 88$\pm$7 K and $\textit{T}_{\rm ex, SUMIRE}$ is 99$\pm$8 K, which are consistent to each other within uncertainty but lower than the assumed value of 120 K. The resulting column density is \textit{N}$\rm _{JPL}$ = (3.1$\pm$0.3) $\times$ 10$^{16}$ cm$^{-2}$ and \textit{N}$\rm _{SUMIRE}$ = (3.8$\pm$0.4) $\times$ 10$^{16}$ cm$^{-2}$.

Lastly, with all the detected \textit{a}-, \textit{b}-, and \textit{c}-type transitions, the column density derived by using JPL data (\textit{N}$\rm _{JPL}$ = (5.7$\pm$0.1) $\times$ 10$^{16}$ cm$^{-2}$) is overestimated by $\sim$10$\%$ compared to SUMIRE data (N$\rm _{SUMIRE}$ = (5.1$\pm$0.1) $\times$ 10$^{16}$ cm$^{-2}$) if the $\textit{T}_{\rm ex}$ is fixed. Without fixing the $\textit{T}_{\rm ex}$, the same column density ((4.3$\pm$0.3) $\times$ 10$^{16}$ cm$^{-2}$) can be derived by using JPL and SUMIRE data but the derived temperature is differed by $\sim$10$\%$. As a result, the CH$_2$DOH/CH$_3$OH ratio obtained in V883 Ori with JPL and SUMIRE data is in the range of (1.3$-$2.8) $\times$ 10$^{-2}$ and (1.6$-$2.3) $\times$ 10$^{-2}$, respectively by adopting \textit{N}(CH$_3$OH) = (2.35$^{+0.65}_{-0.24}$) $\times$ 10$^{18}$ cm $^{-2}$ \citep{Jeong2025}. Hereafter, the column density of (5.1$\pm$0.1) $\times$ 10$^{16}$ cm$^{-2}$ derived by using all transitions, fix $\textit{T}_{\rm ex}$, and SUMIRE data will be used in further discussion.

\begin{figure*}
\includegraphics[width=\textwidth]{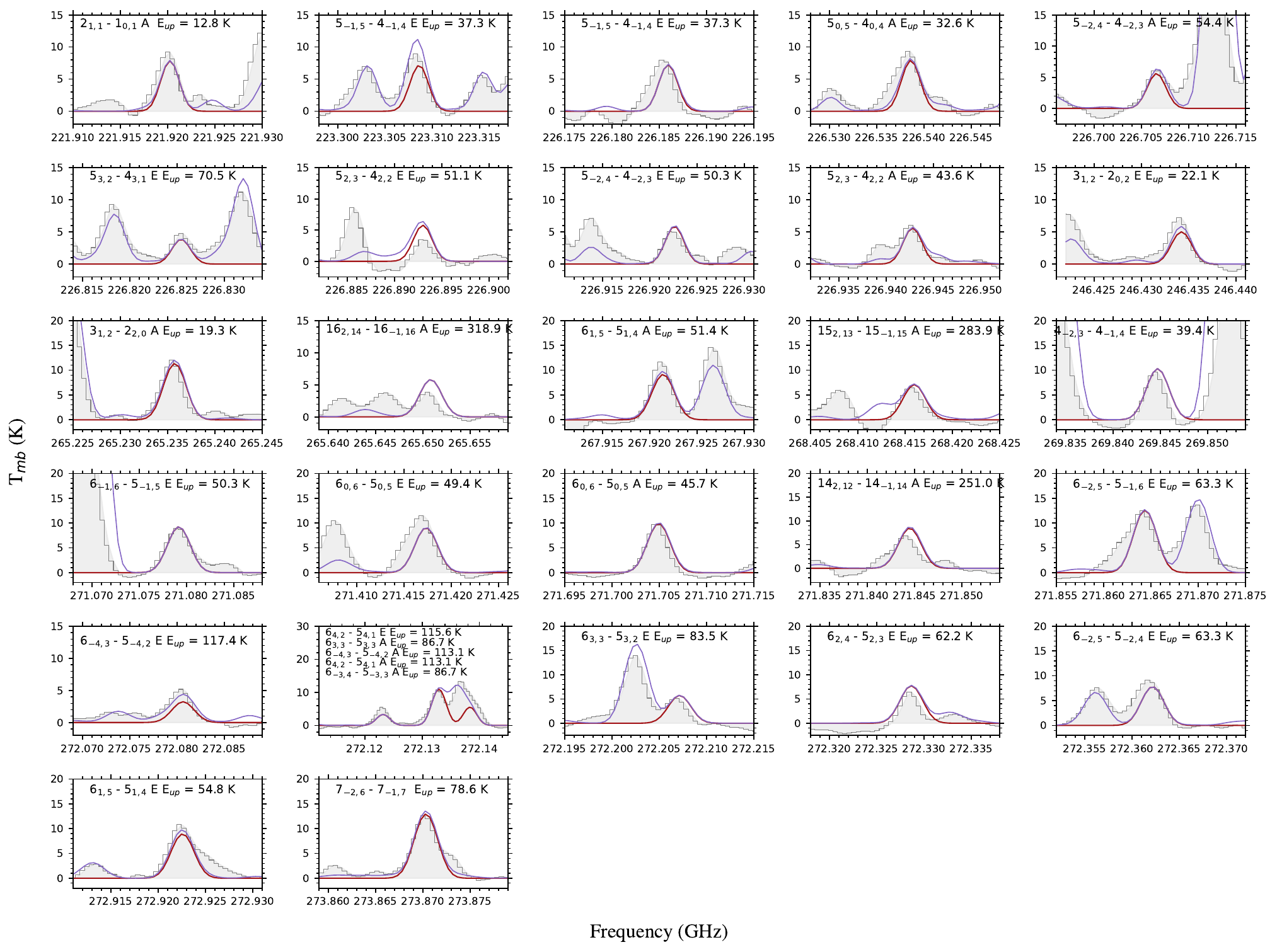}
\centering
\caption{Unblended and optically thin transitions of CH$_3$OD detected in V883 Ori. The observed spectra are plotted as grey histograms in the order of increasing frequency. The best LTE fit by fixing $\textit{T}_{\rm ex}$-120 K is shown in red line and the overall fitting i.e. by including the contribution from all other molecular species detected in V883 Ori in \citet{Jeong2025}, is in purple line. The quantum numbers and value of E$_u$ of each transitions are listed on the upper part of each panel.
\label{fig:CH3OD}}
\end{figure*}


\subsection{\texorpdfstring{CH$_3$OD}{an} and \texorpdfstring{CHD$_2$OH}{lg}}\label{subsec:ch3od}

In the case of CH$_3$OD, 31 unblended and optically thin ($\tau <$0.3) transitions are detected in V883 Ori. The spectroscopic data of unblended CH$_3$OD transitions identified in V883 Ori and the best LTE fitting to the line profile are present in Table \ref{tab:CH3OD_lines} and Figure \ref{fig:CH3OD}, respectively. Fixing the same values of physical parameters as for CH$_2$DOH, the column density of CH$_3$OD is derived to be $\textit{N}$ = (4.22$\pm$0.06) $\times$ 10$^{16}$ cm$^{-2}$ and the corresponding D/H ratio is (1.79$\pm$0.36) $\times$ 10$^{-2}$. By leaving the $\textit{T}_{\rm ex}$ as a free parameter, $\textit{N}$ becomes (3.68$\pm$0.11) $\times$ 10$^{16}$ cm$^{-1}$ and $\textit{T}_{\rm ex}$ is estimated to be 103$\pm$3 K.

In addition, CHD$_2$OH is also searched in V883 Ori by using the catalogue entry (ID: 034506) in the Cologne Database for Molecular Spectroscopy\footnote{\url{https://cdms.astro.uni-koeln.de/classic/}} \citep[CDMS,][]{Endres2016}, which is mainly based on the latest update by \citet{Drozdovskaya2022}. Although CHD$_2$OH is not detected in V883 Ori, an upper limit is estimated to be $\textit{N}$ $\leq$ 5 $\times$ 10$^{15}$ cm$^{-2}$ by assuming the same set of physical parameters as CH$_2$DOH and CH$_3$OD. Taking the CH$_2$DOH column density of  (5.1$\pm$0.1) $\times$ 10$^{16}$ cm$^{-2}$, the upper limit of CHD$_2$OH/CH$_2$DOH ratio is $\sim$0.1.

\begin{deluxetable}{lrlllll}
  \tabletypesize{\scriptsize}
  \tablecolumns{7}
  \tablewidth{0.5\textwidth}
  \tablecaption{Spectroscopic parameters of unblended transitions of CH$_3$OD \citep{Ilyushin2024} observed towards V883 Ori. \label{tab:CH3OD_lines}}
  \tablehead{\colhead{Frequency} & \multicolumn{3}{c}{Transition$^{(a)}$} & \colhead{$E\rm_{up}$} & \colhead{S$_{\mu^2}$} & \colhead{$\log I$} \\
  \colhead{(MHz)} & \colhead{\textit{J$^{\prime}$K$_a^{\prime}$K$_c^{\prime}$}} & \colhead{\textit{J$^{\prime\prime}$K$_a^{\prime\prime}$K$_c^{\prime\prime}$}} & \colhead{E/A} & \colhead{(K)}  & \colhead{(D$^2$)} &  \colhead{(nm$^2$MHz)}}
  \startdata
    \hline
     221920.245 & 2	1	1	& 1	 0	1 &	A &	12.83	& 3.04 & -4.1190 \\
     223308.594 & 5  1   4   & 4  1  3 & A & 38.55   & 3.38 & -4.1048 \\
     226185.940 & 5	-1	5	& 4	-1	4 &	E & 37.26	& 3.35 & -4.0956 \\
     226538.601 & 5	0	5	& 4	 0	4 &	A &	32.63	& 3.54 & -4.0635 \\
     226706.592 & 5	-2	4	& 4	 -2 3 &	A &	54.44	& 2.92 & -4.1781 \\
     226825.424 & 5	3	2	& 4	 3	1 &	E &	70.48	& 2.25 & -4.3140 \\
     226892.984 & 5	2	3	& 4	 2	2 &	E &	51.14	& 2.93 & -4.1711 \\
     226922.611 & 5	-2	4	& 4	 -2 3 &	E &	50.25	& 2.96 & -4.1635 \\
     226942.791 & 5	2	3	& 4	 2	2 &	A &	43.57	& 2.92 & -4.1770 \\
     246434.230 & 3	1	2	& 2	 0	2 &	E &	22.10	& 1.82 & -4.2634 \\
     265235.752 & 3	1	2	& 2	 0	2 &	A &	19.26	& 4.07 & -3.8452 \\
     265650.819 & 16 2   14  & 16 -1 16 &  A & 318.89  & 23.4 & -3.5180 \\
     267921.408 & 6	1	5	& 5	 1	4 &	A &	51.41	& 4.11 & -3.8787 \\
     268415.964 & 15	2	13	& 15 -1	15 & A & 283.85  & 21.5 & -3.4950 \\
     269844.707 & 4	-2	3	& 4	-1	4 &	E &	39.36	& 4.23 & -3.8425 \\
     271079.176 & 6	-1	6	& 5	 -1 5 &	E &	50.27	& 4.07 & -3.8710 \\
     271417.338 & 6	0	6	& 5	0	5 &	E & 49.39	& 3.94 & -3.8827 \\
     271704.941 & 6	0	6	& 5	 0	5 &	A &	45.67	& 4.24 & -3.8445 \\
     271844.484 & 14	2	12	& 14 -1 14 & A & 251.01	& 19.6 & -3.4765 \\
     271864.372 & 6	-2	5	& 6	 -1 6 &	E &	63.32	& 6.48 & -3.6854 \\
     272080.625 & 6	-4	3	& 5	 -4 2 &	E &	117.38	& 2.30 & -4.2128 \\
     272123.196 & 6	4	2	& 5	 4	1 &	E &	115.55	& 2.33 & -4.2044 \\
     272132.638 & 6	3	3	& 5	 3	2 &	A &	86.73	& 3.12 & -4.0358 \\
     272133.154 & 6	-4	3	& 5	 -4 2 &	A &	113.06	& 2.34 & -4.1989 \\
     272133.184 & 6	4	2	& 5	 4	1 &	A &	113.06	& 2.34 & -4.1989 \\
     272138.354 & 6	-3	4	& 5	 -3 3 &	A &	86.73	& 3.12 & -4.0358 \\
     272207.143 & 6	3	3	& 5	 3	2 &	E &	83.54	& 3.17 & -4.0241 \\
     272328.655 & 6	2	4	& 5	 2	3 &	E &	62.21	& 3.73 & -3.9250 \\
     272362.170 & 6	-2	5	& 5	 -2 4 &	E &	63.32	& 3.75 & -3.9213 \\
     272922.578 & 6	1	5	& 5	 1	4 &	E &	54.83	& 4.04 & -3.8749 \\
     273870.319 & 7	-2	6	& 7	 -1 7 &	E &	78.58	& 7.55 & -3.6346 \\
  \enddata
  \tablecomments{$^{(a)}$ Quantum numbers of each CH$_3$OD transition with \textit{J}denotes the angular momentum, \textit{K$_a$} along with + or $-$ designate the projection of \textit{J} on the symmetry axis and the parity, and E/A denotes the transitions of E and A species, respectively.
   }
\end{deluxetable}


\section{Discussion} \label{sec:dis}
\subsection{The Use of Accurately Measured Spectroscopic Data} \label{sec:impact}
\begin{figure}
\centering
\includegraphics[width=\linewidth]{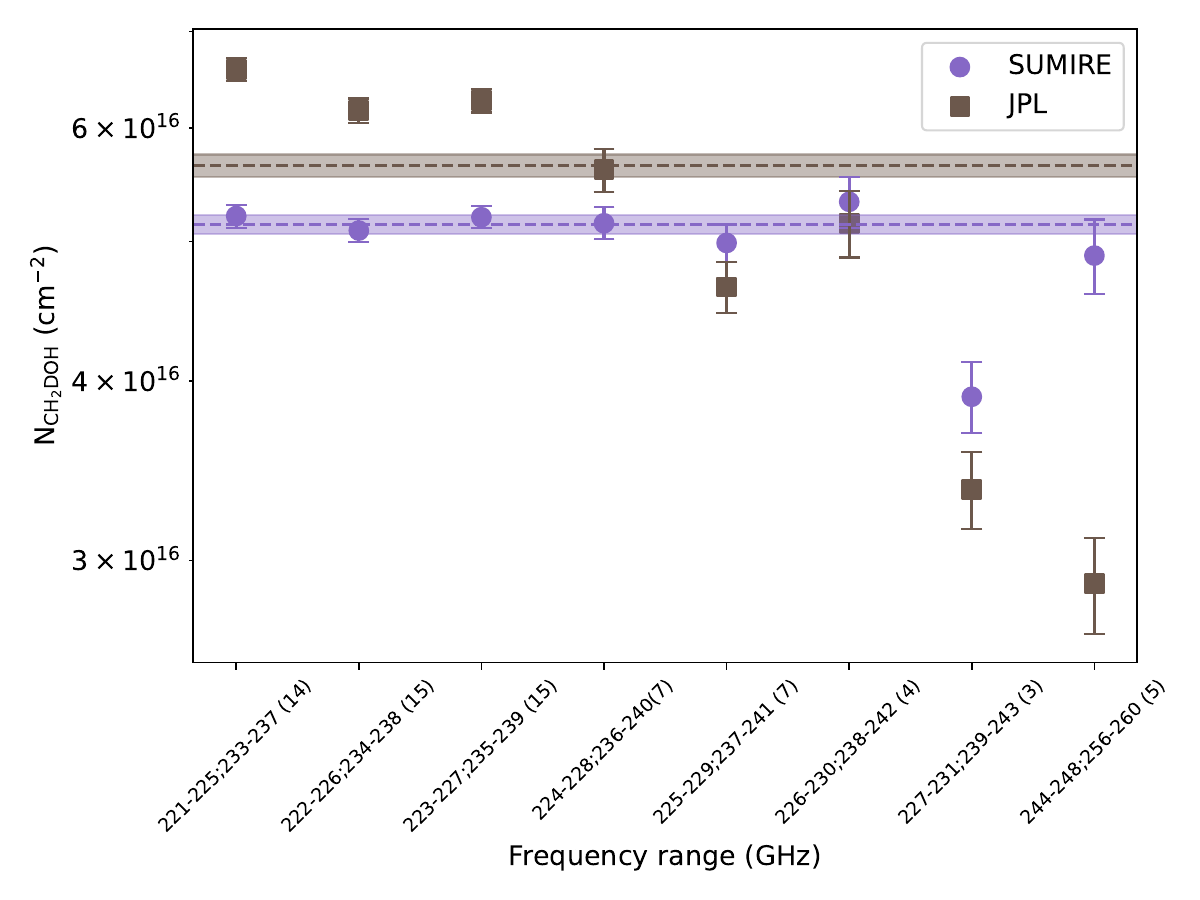} 
\caption{
Derived CH$_2$DOH column density by using different detected transitions at various frequency ranges between 216 and 264 GHz in V883 Ori. The frequency ranges are set by roughly mimicking the spectral windows which are typically used in observation setup \citep[e.g.][]{Fontani2015,Bianchi2017,Ospina-Zamudio2019}. The number in each bracket indicate the number of transitions included in the corresponding frequency range. Results employed SUMIRE and JPL spectroscopic data are indicated in purple and brown respectively. The dash line denotes the CH$_2$DOH column density derived by including all available transitions and the shaded area denotes the associated uncertainty.}
\label{fig:alma_band}
\end{figure}

Spectroscopic data, particularly line frequencies and line intensities listed in molecular spectroscopy databases, are often based on the extrapolation of a limited frequency range observed in laboratory spectroscopic measurements. Parameters for transitions outside the observed range thus tend to exhibit large uncertainties and systematic deviations. The database entry\footnote{\url{https://spec.jpl.nasa.gov/ftp/pub/catalog/doc/d033004.pdf}} also notes that significant errors can arise when determining column densities, especially when using  \textit{b}-type and \textit{c}-type CH$_2$DOH transitions. Furthermore, a recent study by \citet{Oyama2023} revealed that the calculated line intensities listed in the JPL molecular database, even for the most reliable  \textit{a}-type transitions, show notable deviations from SUMIRE laboratory measurements, which were taken directly within the same frequency range. As determined in Section \ref{sec:res}, excitation temperature and column density derived using JPL data generally exhibit a discrepancy of approximately $\sim$10$-20\%$ when compared to SUMIRE data. The column density determined using all transition types shows relatively less discrepancy than when only \textit{a}- or \textit{b}-type transitions are used. This is because the effect of \textit{S}$\mu^2$ deviations is mitigated when multiple transitions of mixed types are included in the analysis. However, in practice, only a limited number of transitions can be detected in the same region due to the restricted spectral window coverage of observations. For instance, Figure \ref{fig:alma_band} shows the CH$_2$DOH column density derived from transitions in various spectral windows typically used in ALMA observational setups. Depending on the frequency range, different numbers and combinations of transitions are included. Assuming the same $\textit{T}_{\rm ex}$, the column density derived in each frequency range using SUMIRE data is markedly consistent, except for one range that includes only three transitions, whereas the column density derived from JPL data can vary by a factor of two. The effect is most pronounced when the frequency range includes only \textit{b}-type and \textit{c}-type transitions, as anticipated from the caution in the database entry. Consequently, previous observational studies of CH$_2$DOH column densities based on a few transitions within the 216 GHz $-$ 264 GHz frequency range are likely to be overestimated or underestimated. This could lead to unreliable determinations of deuterium fractionation in methanol.

It is worth noting that this effect on CH$_2$DOH column density could only be investigated within the 216 GHz$-$264 GHz frequency range in this study due to the limited frequency coverage of currently available laboratory measurements. However, a similar or greater effect is expected, as discrepancies in \textit{S}$\mu^2$ values have already been shown to increase more significantly with higher \textit{J} levels \citep{Oyama2023}. In the case of V883 Ori, CH$_2$DOH has previously been observed with ALMA in Band 3 and 7 i.e. within 84$-$116 GHz and 275$-$373 GHz \citep{Lee2019,Yamato2024}. In ALMA Band 7, five \textit{c}-type transitions of CH$_2$DOH were detected \citep{Lee2019}. Assuming the same excitation temperature of $\textit{T}_{\rm ex}$ = 120 K,  the derived column density is 6.21$^{+0.69}_{-0.74} \times$ 10$^{16}$ cm$^{-2}$ which is a factor of 1.2 higher than the value derived in this work. On the other hand, CH$_2$DOH was tentatively detected through three  \textit{b}-type transitions identified in ALMA Band 3 when considering the molecular emission region of 0.3$^{\prime\prime}$ \citep{Yamato2024}.

Assuming an excitation temperature of 106.8$^{+4.2}_{-3.9}$K derived from CH$_3$OH, \citet{Yamato2024} estimated an upper limit or 2.2$^{+0.4}_{-0.4} \times$ 10$^{16}$ cm$^{-2}$ which is a factor of $\sim$2 lower than the present work by using the same temperature. Given that the column density derived for CH$_3$OH is consistent within uncertainties across different ALMA frequency bands \citep{Lee2019, Yamato2024,Jeong2025}, the observed variation in CH$_2$DOH column density can be explained by deviations in \textit{S}$\mu^2$ values between calculated and directly measured spectroscopic data. Alternatively, the inconsistency may indicate chemical differences, as Band 3 observations probe a more inner region that is optically thick in Bands 6 and 7, potentially resulting in a different CH$_2$DOH abundance. Thus, direct laboratory measurements of CH$_2$DOH spectroscopy over wider frequency ranges or the development of spectroscopic analysis techniques that can provide more accurate intensities and frequencies are urgently needed for accurate determination of CH$_2$DOH column density. 

\subsection{Methanol deuterium fractionation in V883 Ori} \label{sec:v883}
V883 Ori is classified as a young eruptive protostar that is in transition from the Class I to II evolutionary stage. The protostar underwent an accretion outburst, which has increased its luminosity to $\sim$200 $L_{\odot}$ \citep{Furlan2016} and hence elevated the temperature of its well-developed Keplerian-rotating disk \citep{Cieza2016}. Owing to the rapid increase of temperature, a number of molecules including water and COMs can be thermally desorbed into the gas phase which makes it possible to study the chemical composition of the ices in the disk midplane \citep{Hoff2018,Tobin2023,Lee2024,Yamato2024,Jeong2025}. Deuterium fractionation in water has been studied towards the disk of V883 Ori, a disk-averaged HDO/H$_2$O ratio is calculated to be (2.26$\pm$0.63) $\times$ 10$^{-3}$ \citep{Tobin2023}, resulting to a D/H ratio of (1.13$\pm$0.32) $\times$ 10$^{-3}$ after applying the statistical correction. Based on the comparison of the derived HDO/H$_2$O ratio across various sources, including Class 0 protostars, comets, and Earth’s oceans, \citet{Tobin2023} proposed that water in the disk could be directly inherited from the natal star-forming cloud and subsequently incorporated into large icy bodies, like comets, without significant chemical alteration.

\begin{figure}[htbp!]
\centering
\includegraphics[width=0.9\linewidth]{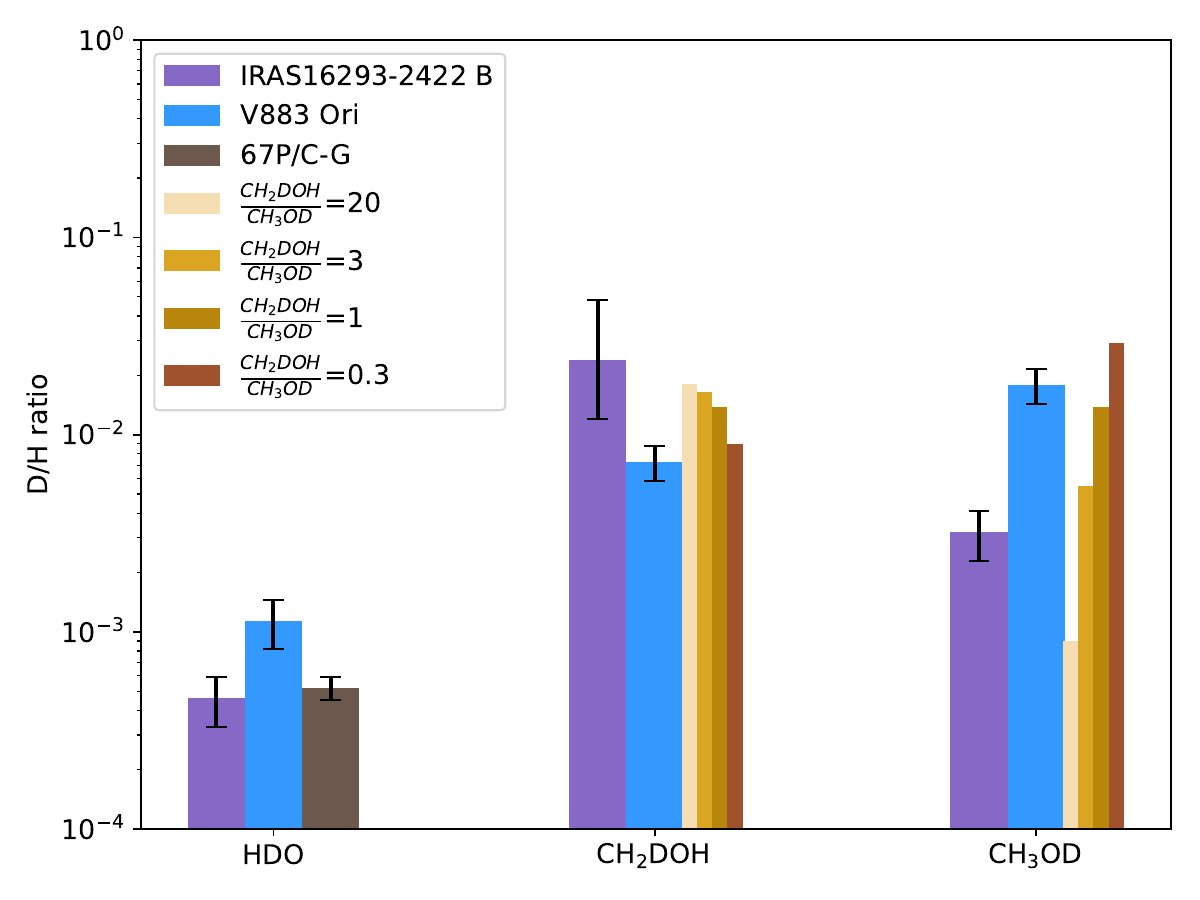} 
\caption{The D/H ratio determined in water and methanol, inferred from CH$_2$DOH, CH$_3$OD, HDO, and CHD$_2$OH, towards the Class 0 protostar IRAS 16293-2422 B (purple) \citep{Persson2013,Jorgensen2018}, the disk around the Class I/II outbursting star V883 Ori (blue) \citep[][, and this work]{Tobin2023}, and the comet 67P/C–G (dark brown) \citep{Drozdovskaya2021}. The D/H ratio is corrected by the relation of CH$_2$DOH/CH$_3$OH = 3(D/H) and HDO/H$_2$O = 2(D/H). For the column density of CH$_2$DOH in IRAS 16293-2422 B, the error bar takes into account a factor of two uncertainty assumed from the JPL spectroscopic data. For CH$_2$DOH and CH$_3$OD in 67P/C–G, beige, light brown, brown, and brick red denote deuteration in methyl ($-$CH$_3$) and hydroxyl($-$OH) considering a $\frac{CH_2DOH}{CH_3OH}$ ratio of 20, 3, 1, and 0.3, respectively.} 
\label{fig:DHratio}
\end{figure}

In this work, the deuterium fractionation in methanol is measured towards the disk of V883 Ori. This allows for a comparative investigation of the chemical heritage of methanol across sources at different evolutionary stages by comparing the D/H ratio determined from CH$_2$DOH, and CH$_3$OD with respect to CH$_3$OH with those observed in the Class 0 protostar IRAS 16293-2422 B (hereafter I16293B) and the comet 67P/Churyumov–Gerasimenko (hereafter 67P/C–G (See Figure \ref{fig:DHratio}). I16293B was selected as the only protostar for comparison in this study for two main reasons. First, it is the only source where CH$_3$OD has been reanalysed using the new spectroscopic data from \citet{Ilyushin2024}, making it the only available study for an accurate determination of CH$_3$OD column density. According to their study, the column density of CH$_3$OD could be significantly affected by incorrect coupling between the partition function and line intensities in previous spectroscopic data. Second, CH$_2$DOH was detected in I16293B through a line survey, meaning that multiple transitions were used to derive its column density, which should be more reliable than relying on only a few transitions, as demonstrated in Section \ref{sec:impact}. Although the I16293B line survey was conducted in ALMA Band 7 (275$-$373 GHz) \citep{Jorgensen2018}, and laboratory-measured line intensities are not yet available for reanalysis, the impact of line intensity, as demonstrated in this study with Band 6 data, is expected to be mitigated when multiple transitions of mixed types are used in the analysis. In Figure \ref{fig:DHratio}, the uncertainty in the CH$_2$DOH column density for I16293B is assumed to be a factor of 2, based on the fact that the largest discrepancy identified in this study is also a factor of 2. However, the actual uncertainty requires further verification through additional analysis with laboratory measurements in the same frequency range.

For CH$_2$DOH, the abundance ratio of CH$_2$DOH/CH$_3$OH in V883 Ori is derived to be (2.2$\pm$0.4) $\times$ 10$^{-2}$, translating to a D/H ratio of (7.3$\pm$1.5) $\times$ 10$^{-3}$ after applying the statistical correction. This is consistent with the average value of D/H ratio derived for low-mass pre-stellar cores ad protostars \citep[(2.2$\pm$1.2) $\times$ 10$^{-2}$;][]{Drozdovskaya2021,vanGelder2022}, without accounting for the effect of \textit{S}$\mu^2$ values. However, when compared to the ratio in I16293B of which the CH$_2$DOH column density is expected to be less affected by spectroscopic data, V883 Ori appears to be lower by a factor of $\sim$3. In comparison to the comet 67P, since CH$_2$DOH and CH$_3$OD cannot be distinguished in the mass spectra from ROSINA, different probable deuteration is assumed for the methyl and hydroxyl group of methanol (Figure \ref{fig:DHratio}). With a CH$_3$OD+CH$_2$DOH to CH$_3$OH ratio of (5.5$\pm$0.5) × 10$^{-2}$ in 67P, the overall variation, regardless of the assumed ratio, remains within a factor of 3. However, the D/H ratio derived for V883 Ori remains the lowest, even when compared to a CH$_2$DOH/CH$_3$OD abundance ratio of 0.3. Conversely, the directly inferred D/H ratio from CH$_3$OD in V883 Ori is 5.6 times higher than in I16293B and ranges from a factor of a few to an order of magnitude higher than in 67P, except when a CH$_2$DOH/CH$_3$OD abundance ratio of 0.3 is assumed.

Compared to the D/H ratio in water studied by \citet{Tobin2023}, the chemical heritage of methanol deuteration remains inconclusive, primarily due to the limited number of sources where the column densities of CH$_2$DOH and CH$_3$OD have been accurately determined using updated spectroscopic data. Increasing the number of studies by applying the refined spectroscopic data to future observations and reanalysing previous studies will provide further insight into the methanol deuteration across various sources at different evolutionary stages.

\subsection{The CH$_2$DOH/CH$_3$OD Ratio} \label{sec:ratio}

By using the SUMIRE spectroscopic data to derive the column density of CH$_2$DOH and CH$_3$OD, a CH$_2$DOH/CH$_3$OD ratio of 1.22$\pm$0.02 is found in the disk around V883 Ori. This contrasts with the protostar I16293B, where the CH$_2$DOH/CH$_3$OD ratio is equal to $\sim$22. Although the CH$_2$DOH column density might be over- or underestimated by using JPL data, this does not alter the finding that the CH$_2$DOH/CH$_3$OD ratio in I16293B is an order of magnitude higher.

As mentioned in Section \ref{sec:intro}, a statistical value of 3 is typically used as the standard CH$_2$DOH/CH$_3$OD ratio for comparison. However, models that consider the formation of both molecules exclusively through barrierless addition reactions at $\sim$10 K predict a ratio close to unity across core ages of $10^3 - 10^6$ years and $10^5 - 10^7$ years in \citet{Taquet2012b} and \citet{Kulterer2022}, respectively. The CH$_2$DOH/CH$_3$OD ratio of 1.22$\pm$0.02 determined in V883 Ori may simply indicate that both CH$_2$DOH and CH$_3$OD form via a single D addition event that replaces one H addition during the hydrogenation of methanol. Notably, CH$_3$OD is believed to form solely through this addition process. The proposed H-abstraction and substitution mechanism, where H is abstracted from methanol followed by D addition or direct H$-$D exchange, has been shown to significantly enhance deuterium fractionation of methanol at the methyl group (forming CH$_2$OD) but not at the hydroxyl group \citep{Nagaoka2005,Nagaoka2007}. Chemical models incorporating this mechanism predict a higher CH$_2$DOH/CH$_3$OD ratio ($\geq$10) \citep{Taquet2012b,Kulterer2022}, which could explain the observed ratio in I16293B. Furthermore, a CH$_2$DOH/CH$_3$OD ratio of unity can also be predicted by models that include abstraction and substitution reactions \citep{Taquet2012b}. A ratio of unity is obtained at short timescale ($\leq$10$^{4}$ yr) whereas 10$-$20 can be obtained at high densities ($\geq10^{6}$ $\rm cm^{-3}$) and longer core evolution times ($\sim5-10\times10^4$ yr). This suggests that V883 Ori may have experienced a shorter core time with relatively lower density in its earlier quiescent phase compared to I16293B, suppressing abstraction and substitution reactions from significantly altering the CH$_2$DOH/CH$_3$OD ratio.

An alternative scenario to explain the CH$_2$DOH/CH$_3$OD ratio of unity in V883 Ori is that CH$_2$DOH might be produced less efficiently, or an additional formation mechanism for CH$_3$OD is at work. Regardless of the formation mechanism, all models in \citet{Kulterer2022} indicate that dust temperature is the most influential factor affecting the CH$_2$DOH/CH$_3$OD ratio. Cores with warmer dust temperature i.e. T=15$-$30 K typically exhibit lower CH$_2$DOH/CH$_3$OD ratios. Through the study of complex organic molecules (COMs), \citet{Yamato2024} proposed that methanol could have been (re-)formed on lukewarm ($\sim$30$–$50 K) dust grain surfaces in the inner region of the disk around V883 Ori during its quiescent phase. This is further supported by the study of \citet{Jeong2025}, which also suggests that COMs forms in warm, water-rich ice environment in this source. Consequently, methanol deuteration is expected to be less efficient than in I16293B, which may explain the lower CH$_2$DOH/CH$_3$OH ratio but not the higher CH$_3$OD/CH$_3$OH ratio derived in V883 Ori. To account for this, additional formation process of CH$_3$OD may need to be invoked at some stage. Rapid H$-$D exchange between non-deuterated methanol and deuterated water i.e. CH$_3$OH + HDO $\rightleftharpoons$ CH$_3$OD + H$_2$O and/or CH$_3$OH + D$_2$O $\rightleftharpoons$ CH$_3$OD + HDO, when the ice reach above 120 K, could serve as one feasible way to enhance CH$_3$OD during warm-up stage. This mechanism has been demonstrated in experimental study \citep{Kawanowa2004} but has not yet been widely considered in modelling work. Whilst only the rate constant for the reverse reaction has been estimated by assuming an activation energy of 4100$\pm$900 K \citep{Faure2015}, a relationship between the forward and backward rate constants has been predicted, as H$-$D exchange reactions are almost thermoneutral. The study found that the CH$_2$DOH/CH$_3$OD ratio scales inversely with the D/H ratio of water, as the hydroxyl group of methanol needs to equilibrate with water ice during the warm-up phase, when the temperature is high enough to trigger the H$-$D exchange. However, the higher D/H ratio of water, which is typically expected at low temperatures and required for the H$–$D exchange, seems to contrast with the proposed warmer dust temperature in V883 Ori during its earlier stage. Given that there are only two data points available for comparison, the constraints on the chemical processes of CH$_2$DOH and CH$_3$OD remain inconclusive.

\section{Conclusions} \label{sec:con}
This works investigated the methanol deuteration in the disk around V883 Ori, based on the recent laboratory measured spectroscopic dataset for CH$_2$DOH \citep{Oyama2023} together with the observational data from the Atacama Large Millimeter/submillimeter Array (ALMA) Spectral Survey of An eruptive Young star \citep[ASSAY;][]{Lee2024}. The main conclusions are as follows:

\begin{enumerate}
  \item Using the more precise spectroscopic dataset measured by SUMIRE \citep{Oyama2023} and published by \citet{Ilyushin2024}, the column density derived for CH$_2$DOH and CH$_3$OD is (5.14$\pm$0.08) $\times $10$^{16}$ cm$^{-2}$ and (4.22$\pm$0.06) $\times$ 10$^{16}$ cm$^{-2}$, respectively. 
  \item Owing to the discrepancy reported in \textit{S}$\mu^2$ values between JPL and SUMIRE spectroscopic data by \citet{Oyama2023}, the effect on \textit{S}$\mu^2$ values on the determination of column density and excitation temperature is investigated. Depending on the selection and number of CH$_2$DOH transitions, results estimated by using SUMIRE data is more consistent whilst a factor of up to two variations can be found by using the calculated \textit{S}$\mu^2$ value listed on JPL database. 
  \item With statistical correction, the D/H ratio of CH$_2$DOH is (7.3$\pm$1.5) $\times$ 10$^{-3}$, which is a factor of $\sim$3 lower than that in Class 0 source IRAS16293-2422 B. On the other hand, the D/H ratio of CH$_3$OD is derived to be (1.79$\pm$0.36) $\times$ 10$^{-2}$. Whilst the D/H ratio from CH$_2$DOH is relatively low compared to Class 0 low-mass protostar and the comet 67P, the D/H ratio derived from CH$_3$OD appears to be higher amongst the three. 
  
  \item The resulting CH$_2$DOH/CH$_3$OD ratio of 1.22$\pm$0.02 not only deviates from the statistical value of 3 but also contrasts with the value determined in I16293B ($\sim$22). It suggests that both CH$_2$DOH and CH$_3$OD form via solely through the addition process in V883 Ori. Alternatively, CH$_2$DOH may form less efficiently in V883 Ori due to the warmer dust temperature in its quiescent phase, as proposed in previous studies. Another possibility is that an additional chemical process, such as H$-$D exchange on hydroxyl group of methanol during warm-up phase, enhances the abundance of CH$_3$OD.  
\end{enumerate}

The analysis illustrates the importance of using laboratory direct measured spectroscopy in order to accurately determine the molecular column density and excitation temperature, which are essential for studies of isotopic fractionation. Concurrently, the works has potentially filled in the gap of investigating the methanol fractionation at different star-forming evolutionary stages. Future studies incorporating updated spectroscopic data, combined with detailed chemical modelling, will provide stringent constraints on methanol deuteration chemistry across various masses and evolutionary stages.

\begin{acknowledgments}
\end{acknowledgments}
S. Z. acknowledge the support by RIKEN Special Postdoctoral Researchers Program. JELee and JHJeong were supported by the NRF grant funded by the Korean government (MSIT) (grant numbers 2021R1A2C1011718 and RS-2024-00416859). Y.-L.Y. acknowledges support from Grant-in-Aid from the Ministry of Education, Culture, Sports, Science, and Technology of Japan (20H05845, 20H05844, 22K20389), and a pioneering project in RIKEN (Evolution of Matter in the Universe). This paper makes use of the following ALMA data: ADS/JAO.ALMA\#2019.1.00377.S. ALMA is a partnership of ESO (representing its member states), NSF (USA) and NINS (Japan), together with NRC (Canada), MOST and ASIAA (Taiwan), and KASI (Republic of Korea), in cooperation with the Republic of Chile. The Joint ALMA Observatory is operated by ESO, AUI/NRAO and NAOJ.
%
\vspace{5mm}
\facilities{ALMA}
\software{Madrid Data Cube Analysis (Madcuba) on ImageJ is a software developed at the Center of Astrobiology (CAB) in Madrid; \citep[https: //cab.inta-csic.es/madcuba/;][version from 2023 August 31.]{Martin2019}}



\appendix

\section{Partition function}
\label{sec:Qrot}

\begin{table}[h]
    \centering
    \caption{Partition function values of CH$_2$DOH at different temperature used in this study.}
    \begin{adjustbox}{width=0.5\textwidth}
    \begin{tabular}{CCC}
        \hline
        \hline      
    T (K) & $Q\rm_{tot (CH_2DOH, JPL)}$ & $Q\rm_{tot (CH_2DOH, SUMIRE)}$ \\
        \hline
        300.000 & 15172.000 & 17238.140   \\
        225.000 & 9410.3473 & 9467.785   \\
        150.000 & 4359.4672 & 4340.315    \\
        75.000  & 1292.7487 & 1280.608    \\
        37.500  & 399.0612  & 393.387     \\
        18.750  & 114.5900  & 112.090    \\
        9.375   & 30.3886   & 29.327     \\
        5.000 & $-$ & 9.055   \\
        2.725 & $-$ & 3.361   \\
        \hline
    \end{tabular}
    \end{adjustbox}
    \label{tab:Qrot}
\end{table}



\bibliography{references.bib}{}

\begin{thebibliography}{}
\expandafter\ifx\csname natexlab\endcsname\relax\def\natexlab#1{#1}\fi
\providecommand{\url}[1]{\href{#1}{#1}}
\providecommand{\dodoi}[1]{doi:~\href{http://doi.org/#1}{\nolinkurl{#1}}}
\providecommand{\doeprint}[1]{\href{http://ascl.net/#1}{\nolinkurl{http://ascl.net/#1}}}
\providecommand{\doarXiv}[1]{\href{https://arxiv.org/abs/#1}{\nolinkurl{https://arxiv.org/abs/#1}}}

\bibitem[{H.~E. {Ambrose} {et~al.}(2021){Ambrose}, {Shirley}, \& {Scibelli}}]{Ambrose2021}
{Ambrose}, H.~E., {Shirley}, Y.~L., \& {Scibelli}, S. 2021, \bibinfo{title}{{A survey of CH$_{2}$DOH towards starless and pre-stellar cores in the Taurus molecular cloud},} \mnras, 501, 347, \dodoi{10.1093/mnras/staa3649}

\bibitem[{A. {Belloche} {et~al.}(2016){Belloche}, {M{\"u}ller}, {Garrod}, \& {Menten}}]{Belloche2016}
{Belloche}, A., {M{\"u}ller}, H.~S.~P., {Garrod}, R.~T., \& {Menten}, K.~M. 2016, \bibinfo{title}{{Exploring molecular complexity with ALMA (EMoCA): Deuterated complex organic molecules in Sagittarius B2(N2)},} \aap, 587, A91, \dodoi{10.1051/0004-6361/201527268}

\bibitem[{E. {Bianchi} {et~al.}(2017){Bianchi}, {Codella}, {Ceccarelli}, {Fontani}, {Testi}, {Bachiller}, {Lefloch}, {Podio}, \& {Taquet}}]{Bianchi2017}
{Bianchi}, E., {Codella}, C., {Ceccarelli}, C., {et~al.} 2017, \bibinfo{title}{{Decrease of the organic deuteration during the evolution of Sun-like protostars: the case of SVS13-A},} \mnras, 467, 3011, \dodoi{10.1093/mnras/stx252}

\bibitem[{E.~G. {B{\o}gelund} {et~al.}(2018){B{\o}gelund}, {McGuire}, {Ligterink}, {Taquet}, {Brogan}, {Hunter}, {Pearson}, {Hogerheijde}, \& {van Dishoeck}}]{Bogelund2018}
{B{\o}gelund}, E.~G., {McGuire}, B.~A., {Ligterink}, N. F.~W., {et~al.} 2018, \bibinfo{title}{{Low levels of methanol deuteration in the high-mass star-forming region NGC 6334I},} \aap, 615, A88, \dodoi{10.1051/0004-6361/201832757}

\bibitem[{P. {Caselli} \& C. {Ceccarelli}(2012){Caselli} \& {Ceccarelli}}]{Caselli2012}
{Caselli}, P., \& {Ceccarelli}, C. 2012, \bibinfo{title}{{Our astrochemical heritage},} \aapr, 20, 56, \dodoi{10.1007/s00159-012-0056-x}

\bibitem[{C. {Ceccarelli} {et~al.}(2014){Ceccarelli}, {Caselli}, {Bockel{\'e}e-Morvan}, {Mousis}, {Pizzarello}, {Robert}, \& {Semenov}}]{Ceccarelli2014}
{Ceccarelli}, C., {Caselli}, P., {Bockel{\'e}e-Morvan}, D., {et~al.} 2014, in Protostars and Planets VI, ed. H.~{Beuther}, R.~S. {Klessen}, C.~P. {Dullemond}, \& T.~{Henning}, 859--882, \dodoi{10.2458/azu_uapress_9780816531240-ch037}

\bibitem[{S.~B. {Charnley} {et~al.}(1997){Charnley}, {Tielens}, \& {Rodgers}}]{Charnley1997}
{Charnley}, S.~B., {Tielens}, A.~G.~G.~M., \& {Rodgers}, S.~D. 1997, \bibinfo{title}{{Deuterated Methanol in the Orion Compact Ridge},} \apjl, 482, L203, \dodoi{10.1086/310697}

\bibitem[{L.~A. {Cieza} {et~al.}(2016){Cieza}, {Casassus}, {Tobin}, {Bos}, {Williams}, {Perez}, {Zhu}, {Caceres}, {Canovas}, {Dunham}, {Hales}, {Prieto}, {Principe}, {Schreiber}, {Ruiz-Rodriguez}, \& {Zurlo}}]{Cieza2016}
{Cieza}, L.~A., {Casassus}, S., {Tobin}, J., {et~al.} 2016, \bibinfo{title}{{Imaging the water snow-line during a protostellar outburst},} \nat, 535, 258, \dodoi{10.1038/nature18612}

\bibitem[{M.~N. {Drozdovskaya} {et~al.}(2022){Drozdovskaya}, {Coudert}, {Margul{\`e}s}, {Coutens}, {J{\o}rgensen}, \& {Manigand}}]{Drozdovskaya2022}
{Drozdovskaya}, M.~N., {Coudert}, L.~H., {Margul{\`e}s}, L., {et~al.} 2022, \bibinfo{title}{{Successive deuteration in low-mass star-forming regions: The case of D$_{2}$-methanol (CHD$_{2}$OH) in IRAS 16293-2422},} \aap, 659, A69, \dodoi{10.1051/0004-6361/202142863}

\bibitem[{M.~N. {Drozdovskaya} {et~al.}(2021){Drozdovskaya}, {Schroeder I}, {Rubin}, {Altwegg}, {van Dishoeck}, {Kulterer}, {De Keyser}, {Fuselier}, \& {Combi}}]{Drozdovskaya2021}
{Drozdovskaya}, M.~N., {Schroeder I}, I. R.~H.~G., {Rubin}, M., {et~al.} 2021, \bibinfo{title}{{Prestellar grain-surface origins of deuterated methanol in comet 67P/Churyumov-Gerasimenko},} \mnras, 500, 4901, \dodoi{10.1093/mnras/staa3387}

\bibitem[{C.~P. {Endres} {et~al.}(2016){Endres}, {Schlemmer}, {Schilke}, {Stutzki}, \& {M{\"u}ller}}]{Endres2016}
{Endres}, C.~P., {Schlemmer}, S., {Schilke}, P., {Stutzki}, J., \& {M{\"u}ller}, H.~S.~P. 2016, \bibinfo{title}{{The Cologne Database for Molecular Spectroscopy, CDMS, in the Virtual Atomic and Molecular Data Centre, VAMDC},} Journal of Molecular Spectroscopy, 327, 95, \dodoi{10.1016/j.jms.2016.03.005}

\bibitem[{A. {Faure} {et~al.}(2015){Faure}, {Faure}, {Theul{\'e}}, {Quirico}, \& {Schmitt}}]{Faure2015}
{Faure}, A., {Faure}, M., {Theul{\'e}}, P., {Quirico}, E., \& {Schmitt}, B. 2015, \bibinfo{title}{{Hydrogen isotope exchanges between water and methanol in interstellar ices},} \aap, 584, A98, \dodoi{10.1051/0004-6361/201526499}

\bibitem[{F. {Fontani} {et~al.}(2015){Fontani}, {Busquet}, {Palau}, {Caselli}, {S{\'a}nchez-Monge}, {Tan}, \& {Audard}}]{Fontani2015}
{Fontani}, F., {Busquet}, G., {Palau}, A., {et~al.} 2015, \bibinfo{title}{{Deuteration and evolution in the massive star formation process. The role of surface chemistry},} \aap, 575, A87, \dodoi{10.1051/0004-6361/201424753}

\bibitem[{G.~W. {Fuchs} {et~al.}(2009){Fuchs}, {Cuppen}, {Ioppolo}, {Romanzin}, {Bisschop}, {Andersson}, {van Dishoeck}, \& {Linnartz}}]{Fuchs2009}
{Fuchs}, G.~W., {Cuppen}, H.~M., {Ioppolo}, S., {et~al.} 2009, \bibinfo{title}{{Hydrogenation reactions in interstellar CO ice analogues. A combined experimental/theoretical approach},} \aap, 505, 629, \dodoi{10.1051/0004-6361/200810784}

\bibitem[{E. {Furlan} {et~al.}(2016){Furlan}, {Fischer}, {Ali}, {Stutz}, {Stanke}, {Tobin}, {Megeath}, {Osorio}, {Hartmann}, {Calvet}, {Poteet}, {Booker}, {Manoj}, {Watson}, \& {Allen}}]{Furlan2016}
{Furlan}, E., {Fischer}, W.~J., {Ali}, B., {et~al.} 2016, \bibinfo{title}{{The Herschel Orion Protostar Survey: Spectral Energy Distributions and Fits Using a Grid of Protostellar Models},} \apjs, 224, 5, \dodoi{10.3847/0067-0049/224/1/5}

\bibitem[{R. {Garrod} {et~al.}(2006){Garrod}, {Park}, {Caselli}, \& {Herbst}}]{Garrod2006}
{Garrod}, R., {Park}, I.~H., {Caselli}, P., \& {Herbst}, E. 2006, \bibinfo{title}{{Are gas-phase models of interstellar chemistry tenable? The case of methanol},} Faraday Discussions, 133, 51, \dodoi{10.1039/b516202e}

\bibitem[{W.~D. {Geppert} {et~al.}(2005){Geppert}, {Hellberg}, {{\"O}sterdahl}, {Semaniak}, {Millar}, {Roberts}, {Thomas}, {Hamberg}, {Ugglas}, {Ehlerding}, {Zhaunerchyk}, {Kaminska}, \& {Larsson}}]{Geppert2005}
{Geppert}, W.~D., {Hellberg}, F., {{\"O}sterdahl}, F., {et~al.} 2005, in IAU Symposium, Vol. 231, Astrochemistry: Recent Successes and Current Challenges, ed. D.~C. {Lis}, G.~A. {Blake}, \& E.~{Herbst}, 117--124, \dodoi{10.1017/S1743921306007101}

\bibitem[{V.~V. {Ilyushin} {et~al.}(2024){Ilyushin}, {M{\"u}ller}, {Drozdovskaya}, {J{\o}rgensen}, {Bauerecker}, {Maul}, {Porohovoi}, {Alekseev}, {Dorovskaya}, {Zakharenko}, {Lewen}, {Schlemmer}, {Xu}, \& {Lees}}]{Ilyushin2024}
{Ilyushin}, V.~V., {M{\"u}ller}, H.~S.~P., {Drozdovskaya}, M.~N., {et~al.} 2024, \bibinfo{title}{{Rotational spectroscopy of CH$_{3}$OD with a reanalysis of CH$_{3}$OD toward IRAS 16293-2422},} \aap, 687, A220, \dodoi{10.1051/0004-6361/202449918}

\bibitem[{J.-H. {Jeong} {et~al.}(2025){Jeong}, {Lee}, {Lee}, {Baek}, {Kang}, {Lee}, {Kim}, {Yun}, {Aikawa}, {Herczeg}, {Johnstone}, \& {Cieza}}]{Jeong2025}
{Jeong}, J.-H., {Lee}, J.-E., {Lee}, S., {et~al.} 2025, \bibinfo{title}{{ALMA Spectral Survey of an Eruptive Young Star, V883 Ori (ASSAY). II. Freshly Sublimated Complex Organic Molecules in the Keplerian Disk},} \apjs, 276, 49, \dodoi{10.3847/1538-4365/ad9450}

\bibitem[{J.~K. {J{\o}rgensen} {et~al.}(2018){J{\o}rgensen}, {M{\"u}ller}, {Calcutt}, {Coutens}, {Drozdovskaya}, {{\"O}berg}, {Persson}, {Taquet}, {van Dishoeck}, \& {Wampfler}}]{Jorgensen2018}
{J{\o}rgensen}, J.~K., {M{\"u}ller}, H.~S.~P., {Calcutt}, H., {et~al.} 2018, \bibinfo{title}{{The ALMA-PILS survey: isotopic composition of oxygen-containing complex organic molecules toward IRAS 16293-2422B},} \aap, 620, A170, \dodoi{10.1051/0004-6361/201731667}

\bibitem[{H. {Kawanowa} {et~al.}(2004){Kawanowa}, {Kondo}, {Gotoh}, \& {Souda}}]{Kawanowa2004}
{Kawanowa}, H., {Kondo}, M., {Gotoh}, Y., \& {Souda}, R. 2004, \bibinfo{title}{{Hydration and H/D exchange of CH $_{3}$OH adsorbed on the D $_{2}$O-ice surface studied by time-of-flight secondary-ion mass spectrometry (TOF-SIMS)},} Surface Science, 566-568, 1190, \dodoi{10.1016/j.susc.2004.06.086}

\bibitem[{B.~M. {Kulterer} {et~al.}(2022){Kulterer}, {Drozdovskaya}, {Antonellini}, {Walsh}, \& {Millar}}]{Kulterer2022}
{Kulterer}, B.~M., {Drozdovskaya}, M.~N., {Antonellini}, S., {Walsh}, C., \& {Millar}, T.~J. 2022, \bibinfo{title}{{Fevering Interstellar Ices Have More CH3OD},} ACS Earth and Space Chemistry, 6, 1171, \dodoi{10.1021/acsearthspacechem.1c00340}

\bibitem[{J.-E. {Lee} \& E.~A. {Bergin}(2015){Lee} \& {Bergin}}]{Lee2015}
{Lee}, J.-E., \& {Bergin}, E.~A. 2015, \bibinfo{title}{{The D/H Ratio of Water Ice at Low Temperatures},} \apj, 799, 104, \dodoi{10.1088/0004-637X/799/1/104}

\bibitem[{J.-E. {Lee} {et~al.}(2019){Lee}, {Lee}, {Baek}, {Aikawa}, {Cieza}, {Yoon}, {Herczeg}, {Johnstone}, \& {Casassus}}]{Lee2019}
{Lee}, J.-E., {Lee}, S., {Baek}, G., {et~al.} 2019, \bibinfo{title}{{The ice composition in the disk around V883 Ori revealed by its stellar outburst},} Nature Astronomy, 3, 314, \dodoi{10.1038/s41550-018-0680-0}

\bibitem[{J.-E. {Lee} {et~al.}(2024){Lee}, {Kim}, {Lee}, {Lee}, {Baek}, {Yun}, {Aikawa}, {Johnstone}, {Herczeg}, \& {Cieza}}]{Lee2024}
{Lee}, J.-E., {Kim}, C.-H., {Lee}, S., {et~al.} 2024, \bibinfo{title}{{ALMA Spectral Survey of an Eruptive Young Star, V883 Ori (ASSAY). I. What Triggered the Current Episode of Eruption?},} \apj, 966, 119, \dodoi{10.3847/1538-4357/ad3106}

\bibitem[{J.~L. {Linsky}(2003){Linsky}}]{Linsky2003}
{Linsky}, J.~L. 2003, \bibinfo{title}{{Atomic Deuterium/Hydrogen in the Galaxy},} \ssr, 106, 49, \dodoi{10.1023/A:1024673217736}

\bibitem[{S. {Mart{\'\i}n} {et~al.}(2019){Mart{\'\i}n}, {Mart{\'\i}n-Pintado}, {Blanco-S{\'a}nchez}, {Rivilla}, {Rodr{\'\i}guez-Franco}, \& {Rico-Villas}}]{Martin2019}
{Mart{\'\i}n}, S., {Mart{\'\i}n-Pintado}, J., {Blanco-S{\'a}nchez}, C., {et~al.} 2019, \bibinfo{title}{{Spectral Line Identification and Modelling (SLIM) in the MAdrid Data CUBe Analysis (MADCUBA) package. Interactive software for data cube analysis},} \aap, 631, A159, \dodoi{10.1051/0004-6361/201936144}

\bibitem[{I. {Mukhopadhyay}(1997){Mukhopadhyay}}]{Mukhopadhyay1997}
{Mukhopadhyay}, I. 1997, \bibinfo{title}{{Torsional energies, matrix elements and relative intensities of far-infrared absorption transitions in CH $_{2}$DOH},} Spectrochimica Acta Part A: Molecular Spectroscopy, 53, 1947, \dodoi{10.1016/S1386-1425(97)00082-6}

\bibitem[{A. {Nagaoka} {et~al.}(2005){Nagaoka}, {Watanabe}, \& {Kouchi}}]{Nagaoka2005}
{Nagaoka}, A., {Watanabe}, N., \& {Kouchi}, A. 2005, \bibinfo{title}{{H-D Substitution in Interstellar Solid Methanol: A Key Route for D Enrichment},} \apjl, 624, L29, \dodoi{10.1086/430304}

\bibitem[{A. {Nagaoka} {et~al.}(2007){Nagaoka}, {Watanabe}, \& {Kouchi}}]{Nagaoka2007}
{Nagaoka}, A., {Watanabe}, N., \& {Kouchi}, A. 2007, \bibinfo{title}{{Effective Rate Constants for the Surface Reaction between Solid Methanol and Deuterium Atoms at 10 K},} Journal of Physical Chemistry A, 111, 3016, \dodoi{10.1021/jp068978r}

\bibitem[{H. {Nomura} {et~al.}(2022){Nomura}, {Furuya}, {Cordiner}, {Charnley}, {Alexander}, {Nixon}, {Guzman}, {Yurimoto}, {Tsukagoshi}, \& {Iino}}]{Nomura2022}
{Nomura}, H., {Furuya}, K., {Cordiner}, M.~A., {et~al.} 2022, \bibinfo{title}{{The Isotopic Links from Planet Forming Regions to the Solar System},} arXiv e-prints, arXiv:2203.10863, \dodoi{10.48550/arXiv.2203.10863}

\bibitem[{Y. {Osamura} {et~al.}(2004){Osamura}, {Roberts}, \& {Herbst}}]{Osamura2004}
{Osamura}, Y., {Roberts}, H., \& {Herbst}, E. 2004, \bibinfo{title}{{On the possible interconversion between pairs of deuterated isotopomers of methanol, its ion, and its protonated ion in star-forming regions},} \aap, 421, 1101, \dodoi{10.1051/0004-6361:20035762}

\bibitem[{J. {Ospina-Zamudio} {et~al.}(2019){Ospina-Zamudio}, {Favre}, {Kounkel}, {Xu}, {Neill}, {Lefloch}, {Faure}, {Bergin}, {Fedele}, \& {Hartmann}}]{Ospina-Zamudio2019}
{Ospina-Zamudio}, J., {Favre}, C., {Kounkel}, M., {et~al.} 2019, \bibinfo{title}{{Deuterated methanol toward NGC 7538-IRS1},} \aap, 627, A80, \dodoi{10.1051/0004-6361/201834948}

\bibitem[{T. {Oyama} {et~al.}(2023){Oyama}, {Ohno}, {Tamanai}, {Watanabe}, {Yamamoto}, {Sakai}, {Zeng}, {Nakatani}, \& {Sakai}}]{Oyama2023}
{Oyama}, T., {Ohno}, Y., {Tamanai}, A., {et~al.} 2023, \bibinfo{title}{{Laboratory Measurement of CH$_{2}$DOH Line Intensities in the Millimeter-wave Region},} \apj, 957, 4, \dodoi{10.3847/1538-4357/acf320}

\bibitem[{B. {Parise} {et~al.}(2006){Parise}, {Ceccarelli}, {Tielens}, {Castets}, {Caux}, {Lefloch}, \& {Maret}}]{Parise2006}
{Parise}, B., {Ceccarelli}, C., {Tielens}, A.~G.~G.~M., {et~al.} 2006, \bibinfo{title}{{Testing grain surface chemistry: a survey of deuterated formaldehyde and methanol in low-mass class 0 protostars},} \aap, 453, 949, \dodoi{10.1051/0004-6361:20054476}

\bibitem[{B. {Parise} {et~al.}(2002){Parise}, {Ceccarelli}, {Tielens}, {Herbst}, {Lefloch}, {Caux}, {Castets}, {Mukhopadhyay}, {Pagani}, \& {Loinard}}]{Parise2002}
{Parise}, B., {Ceccarelli}, C., {Tielens}, A.~G.~G.~M., {et~al.} 2002, \bibinfo{title}{{Detection of doubly-deuterated methanol in the solar-type protostar IRAS 16293-2422},} \aap, 393, L49, \dodoi{10.1051/0004-6361:20021131}

\bibitem[{J.~C. {Pearson} {et~al.}(2012){Pearson}, {Yu}, \& {Drouin}}]{Pearson2012}
{Pearson}, J.~C., {Yu}, S., \& {Drouin}, B.~J. 2012, \bibinfo{title}{{The ground state torsion rotation spectrum of CH$_{2}$DOH},} Journal of Molecular Spectroscopy, 280, 119, \dodoi{10.1016/j.jms.2012.06.012}

\bibitem[{M.~V. {Persson} {et~al.}(2013){Persson}, {J{\o}rgensen}, \& {van Dishoeck}}]{Persson2013}
{Persson}, M.~V., {J{\o}rgensen}, J.~K., \& {van Dishoeck}, E.~F. 2013, \bibinfo{title}{{Warm water deuterium fractionation in IRAS 16293-2422. The high-resolution ALMA and SMA view},} \aap, 549, L3, \dodoi{10.1051/0004-6361/201220638}

\bibitem[{H.~M. {Pickett} {et~al.}(1998){Pickett}, {Poynter}, {Cohen}, {Delitsky}, {Pearson}, \& {M{\"u}ller}}]{Pickett1998}
{Pickett}, H.~M., {Poynter}, R.~L., {Cohen}, E.~A., {et~al.} 1998, \bibinfo{title}{{Submillimeter, millimeter and microwave spectral line catalog.},} \jqsrt, 60, 883, \dodoi{10.1016/S0022-4073(98)00091-0}

\bibitem[{T. {Prodanovi{\'c}} {et~al.}(2010){Prodanovi{\'c}}, {Steigman}, \& {Fields}}]{Prodanovic2010}
{Prodanovi{\'c}}, T., {Steigman}, G., \& {Fields}, B.~D. 2010, \bibinfo{title}{{The deuterium abundance in the local interstellar medium},} \mnras, 406, 1108, \dodoi{10.1111/j.1365-2966.2010.16734.x}

\bibitem[{A. {Ratajczak} {et~al.}(2009){Ratajczak}, {Quirico}, {Faure}, {Schmitt}, \& {Ceccarelli}}]{Ratajczak2009}
{Ratajczak}, A., {Quirico}, E., {Faure}, A., {Schmitt}, B., \& {Ceccarelli}, C. 2009, \bibinfo{title}{{Hydrogen/deuterium exchange in interstellar ice analogs},} \aap, 496, L21, \dodoi{10.1051/0004-6361/200911679}

\bibitem[{A. {Ratajczak} {et~al.}(2011){Ratajczak}, {Taquet}, {Kahane}, {Ceccarelli}, {Faure}, \& {Quirico}}]{Ratajczak2011}
{Ratajczak}, A., {Taquet}, V., {Kahane}, C., {et~al.} 2011, \bibinfo{title}{{The puzzling deuteration of methanol in low- to high-mass protostars},} \aap, 528, L13, \dodoi{10.1051/0004-6361/201016402}

\bibitem[{J.~C. {Santos} {et~al.}(2022){Santos}, {Chuang}, {Lamberts}, {Fedoseev}, {Ioppolo}, \& {Linnartz}}]{Santos2022}
{Santos}, J.~C., {Chuang}, K.-J., {Lamberts}, T., {et~al.} 2022, \bibinfo{title}{{First Experimental Confirmation of the CH$_{3}$O + H$_{2}$CO {\textrightarrow} CH$_{3}$OH + HCO Reaction: Expanding the CH$_{3}$OH Formation Mechanism in Interstellar Ices},} \apjl, 931, L33, \dodoi{10.3847/2041-8213/ac7158}

\bibitem[{A. {Serrallach} {et~al.}(1974){Serrallach}, {Meyer}, \& {G{\"u}nthard}}]{Serrallach1974}
{Serrallach}, A., {Meyer}, R., \& {G{\"u}nthard}, H.~H. 1974, \bibinfo{title}{{Methanol and deuterated species: Infrared data, valence force field, rotamers, and conformation},} Journal of Molecular Spectroscopy, 52, 94, \dodoi{10.1016/0022-2852(74)90008-3}

\bibitem[{M.~A.~J. {Simons} {et~al.}(2020){Simons}, {Lamberts}, \& {Cuppen}}]{Simons2020}
{Simons}, M.~A.~J., {Lamberts}, T., \& {Cuppen}, H.~M. 2020, \bibinfo{title}{{Formation of COMs through CO hydrogenation on interstellar grains},} \aap, 634, A52, \dodoi{10.1051/0004-6361/201936522}

\bibitem[{V. {Taquet} {et~al.}(2012{\natexlab{a}}){Taquet}, {Ceccarelli}, \& {Kahane}}]{Taquet2012a}
{Taquet}, V., {Ceccarelli}, C., \& {Kahane}, C. 2012{\natexlab{a}}, \bibinfo{title}{{Multilayer modeling of porous grain surface chemistry. I. The GRAINOBLE model},} \aap, 538, A42, \dodoi{10.1051/0004-6361/201117802}

\bibitem[{V. {Taquet} {et~al.}(2012{\natexlab{b}}){Taquet}, {Ceccarelli}, \& {Kahane}}]{Taquet2012b}
{Taquet}, V., {Ceccarelli}, C., \& {Kahane}, C. 2012{\natexlab{b}}, \bibinfo{title}{{Formaldehyde and Methanol Deuteration in Protostars: Fossils from a Past Fast High-density Pre-collapse Phase},} \apjl, 748, L3, \dodoi{10.1088/2041-8205/748/1/L3}

\bibitem[{V. {Taquet} {et~al.}(2014){Taquet}, {Charnley}, \& {Sipil{\"a}}}]{Taquet2014}
{Taquet}, V., {Charnley}, S.~B., \& {Sipil{\"a}}, O. 2014, \bibinfo{title}{{Multilayer Formation and Evaporation of Deuterated Ices in Prestellar and Protostellar Cores},} \apj, 791, 1, \dodoi{10.1088/0004-637X/791/1/1}

\bibitem[{V. {Taquet} {et~al.}(2013){Taquet}, {Peters}, {Kahane}, {Ceccarelli}, {L{\'o}pez-Sepulcre}, {Toubin}, {Duflot}, \& {Wiesenfeld}}]{Taquet2013}
{Taquet}, V., {Peters}, P.~S., {Kahane}, C., {et~al.} 2013, \bibinfo{title}{{Water ice deuteration: a tracer of the chemical history of protostars},} \aap, 550, A127, \dodoi{10.1051/0004-6361/201220084}

\bibitem[{V. {Taquet} {et~al.}(2019){Taquet}, {Bianchi}, {Codella}, {Persson}, {Ceccarelli}, {Cabrit}, {J{\o}rgensen}, {Kahane}, {L{\'o}pez-Sepulcre}, \& {Neri}}]{Taquet2019}
{Taquet}, V., {Bianchi}, E., {Codella}, C., {et~al.} 2019, \bibinfo{title}{{Interferometric observations of warm deuterated methanol in the inner regions of low-mass protostars},} \aap, 632, A19, \dodoi{10.1051/0004-6361/201936044}

\bibitem[{J.~J. {Tobin} {et~al.}(2023){Tobin}, {van't Hoff}, {Leemker}, {van Dishoeck}, {Paneque-Carre{\~n}o}, {Furuya}, {Harsono}, {Persson}, {Cleeves}, {Sheehan}, \& {Cieza}}]{Tobin2023}
{Tobin}, J.~J., {van't Hoff}, M. L.~R., {Leemker}, M., {et~al.} 2023, \bibinfo{title}{{Deuterium-enriched water ties planet-forming disks to comets and protostars},} \nat, 615, 227, \dodoi{10.1038/s41586-022-05676-z}

\bibitem[{M.~L. {van Gelder} {et~al.}(2022){van Gelder}, {Jaspers}, {Nazari}, {Ahmadi}, {van Dishoeck}, {Beltr{\'a}n}, {Fuller}, {S{\'a}nchez-Monge}, \& {Schilke}}]{vanGelder2022}
{van Gelder}, M.~L., {Jaspers}, J., {Nazari}, P., {et~al.} 2022, \bibinfo{title}{{Methanol deuteration in high-mass protostars},} \aap, 667, A136, \dodoi{10.1051/0004-6361/202244471}

\bibitem[{M.~L.~R. {van 't Hoff} {et~al.}(2018){van 't Hoff}, {Tobin}, {Trapman}, {Harsono}, {Sheehan}, {Fischer}, {Megeath}, \& {van Dishoeck}}]{Hoff2018}
{van 't Hoff}, M. L.~R., {Tobin}, J.~J., {Trapman}, L., {et~al.} 2018, \bibinfo{title}{{Methanol and its Relation to the Water Snowline in the Disk around the Young Outbursting Star V883 Ori},} \apjl, 864, L23, \dodoi{10.3847/2041-8213/aadb8a}

\bibitem[{N. {Watanabe} \& A. {Kouchi}(2002){Watanabe} \& {Kouchi}}]{Watanabe2002}
{Watanabe}, N., \& {Kouchi}, A. 2002, \bibinfo{title}{{Efficient Formation of Formaldehyde and Methanol by the Addition of Hydrogen Atoms to CO in H$_{2}$O-CO Ice at 10 K},} \apjl, 571, L173, \dodoi{10.1086/341412}

\bibitem[{Y. {Watanabe} {et~al.}(2021){Watanabe}, {Chiba}, {Sakai}, {Tamanai}, {Suzuki}, \& {Sakai}}]{Watanabe2021}
{Watanabe}, Y., {Chiba}, Y., {Sakai}, T., {et~al.} 2021, \bibinfo{title}{{Spectrometer Using superconductor MIxer Receiver (SUMIRE) for laboratory submillimeter spectroscopy},} \pasj, 73, 372, \dodoi{10.1093/pasj/psab005}

\bibitem[{O.~H. {Wilkins} \& G.~A. {Blake}(2022){Wilkins} \& {Blake}}]{Wilkins2022}
{Wilkins}, O.~H., \& {Blake}, G.~A. 2022, \bibinfo{title}{{Relationship between CH3OD Abundance and Temperature in the Orion KL Nebula},} Journal of Physical Chemistry A, 126, 6473, \dodoi{10.1021/acs.jpca.2c01309}

\bibitem[{Y. {Yamato} {et~al.}(2024){Yamato}, {Notsu}, {Aikawa}, {Okoda}, {Nomura}, \& {Sakai}}]{Yamato2024}
{Yamato}, Y., {Notsu}, S., {Aikawa}, Y., {et~al.} 2024, \bibinfo{title}{{Chemistry of Complex Organic Molecules in the V883 Ori Disk Revealed by ALMA Band 3 Observations},} \aj, 167, 66, \dodoi{10.3847/1538-3881/ad11d9}

\bibitem[{H.-S. {Yun} \& J.-E. {Lee}(2023{\natexlab{a}}){Yun} \& {Lee}}]{Yun2023}
{Yun}, H.-S., \& {Lee}, J.-E. 2023{\natexlab{a}}, \bibinfo{title}{{The Principal Component Analysis Filtering Method for an Unbiased Spectral Survey of Complex Organic Molecules},} \apj, 958, 113, \dodoi{10.3847/1538-4357/acfa6a}

\bibitem[{H.-S. {Yun} \& J.-E. {Lee}(2023{\natexlab{b}}){Yun} \& {Lee}}]{PCA_2023}
{Yun}, H.-S., \& {Lee}, J.-E. 2023{\natexlab{b}}, \bibinfo{title}{{The Principal Component Analysis Filtering Method for an Unbiased Spectral Survey of Complex Organic Molecules},} \apj, 958, 113, \dodoi{10.3847/1538-4357/acfa6a}

\end{thebibliography}
\bibliographystyle{aasjournal}



\end{document}